\documentclass[12pt,superscriptaddress,nofootinbib,aps,floatfix,amsfonts,preprintnumbers,amsmath,amssymb,groupedaddress]{revtex4} 

\usepackage{amsmath}
\usepackage{amsmath}
\usepackage{bbm}
\usepackage{float}
\usepackage{graphicx}	
\usepackage{hyperref}
\usepackage{mathtools}
\usepackage{cancel}
\usepackage{color}

\usepackage[normalem]{ulem}

\usepackage{multirow}
\usepackage{rotating}
\usepackage{subcaption}

\def\lsim{\raise0.3ex\hbox{$\;<$\kern-0.75em\raise-1.1ex
\hbox{$\sim\;$}}}
\def\gsim{\raise0.3ex\hbox{$\;>$\kern-0.75em\raise-1.1ex
\hbox{$\sim\;$}}}

\newcommand{\orcid}[1]{\href{https://orcid.org/#1}{#1}}

\begin{document}

\title{\vspace*{1cm} 
Neutrino Oscillations in Matter using the\\
Adjugate of the Hamiltonian}

\author{Asli Abdullahi} 
\author{Stephen J. Parke}
\email{asli@fnal.gov}
\email{parke@fnal.gov}

\affiliation{Particle Theory Dept., Fermi National Accelerator Laboratory, Batavia, IL, USA}

\begin{abstract} 
\vspace*{5mm}
We revisit neutrino oscillations in constant matter density for a number of different scenarios:
three flavors with the standard Wolfenstein matter potential, four flavors with standard matter potential and three flavors with non-standard matter potentials. 
 To calculate the oscillation probabilities for these scenarios one must determine the eigenvalues and eigenvectors of the Hamiltonians.  We use a method for calculating the eigenvalues that is well known, determination of the zeros of determinant of matrix 
$(\lambda I -H)$, where H is the Hamiltonian, I the identity matrix and $\lambda$ is a scalar. To calculate the associated eigenvectors we use a method that is little known in the particle physics community, the calculation of the adjugate (transpose of the cofactor matrix) of the same matrix, $(\lambda I -H)$.  This method can be applied to any Hamiltonian, but provides a very simple way to determine the eigenvectors for neutrino oscillation in matter, independent of the complexity of the matter potential. This method can be trivially automated using the Faddeev-LeVerrier algorithm for numerical calculations.  For the above scenarios we derive a number of quantities that are invariant of the matter potential, many are new such as the generalization of the Naumov-Harrison-Scott identity for four or more flavors of neutrinos.  We also show how these matter potential independent quantities become matter potential dependent when off-diagonal non-standard matter effects are included.
\end{abstract}
\preprint{FERMILAB-Pub-22-922-T}
\date{August 11, 2023}
\maketitle
\orcid{orcid AA: 0000-0002-6122-4986} \quad 
\orcid{orcid SJP: 0000-0003-2028-6782}

\newpage

\section{Introduction}

The Wolfenstein matter effect \cite{Wolfenstein:1977ue} has been and continues to be very important for exploring neutrino mixing and masses. It is the solution to the solar neutrino anomaly\cite{Mikheev:1986gs, Parke:1986jy}, discovered by Davis et. al. \cite{Cleveland:1994er} and in particular the SNO experiment  \cite{SNO:2002tuh}  determined $|U_{e2}|^2$ ($\sin^2 \theta_{12} \approx 0.3$) and that the sign\footnote{Our labelling of the mass eigenstates is determined by 
$|U_{e1}|^2 > |U_{e2}|^2 > |U_{e3}|^2$ and the form of the Pontecorvo, Maki, Nakagawa, Sakata (PMNS) matrix, U, as in the PDG \cite{ParticleDataGroup:2022pth}. } of $\Delta m^2_{21} $ was positive , i.e. that the state with most $\nu_e$, $\nu_1$, is lighter than state with the next most $\nu_e$, $\nu_2$. 
The current long baseline  (LBL) experiments, T2K \cite{T2K:2011qtm} and NOvA \cite{NOvA:2004blv}, as well as the future experiments, Hyper-Kamiokande (HK) \cite{Hyper-Kamiokande:2016srs}, Korean Neutrino observatory (KNO) \cite{Hyper-Kamiokande:2016srs} and DUNE \cite{DUNE:2015lol}, will determine the parameters $|U_{\mu 3}|^2$ ($\approx  \sin^2 \theta_{23}$)
and $\Delta m^2_{32}$ including its sign.   If $|U_{\mu 3}|^2 >1/2$ ($|U_{\mu 3}|^2 < 1/2$) then 
$\nu_\mu$ ($\nu_\tau$)  dominates the neutrino mass eigenstate with the least $\nu_e$, $\nu_3$. The sign of $\Delta m^2_{32}$ determines the atmospheric mass ordering, whether $\nu_3$ is the heaviest or lightest mass state in the neutrino spectrum. Most importantly, these experiments will determine whether there is significant CP violation in the neutrino sector via the determination of the CP violating  Jarlskog invariant \cite{Jarlskog:1985ht} (or the phase $\delta_{CP}$).  \\

None of these LBL experiments are performed in vacuum as the neutrino beams passes through 295 to 1300 km of the earth's crust. Therefore the Wolfenstein matter effect must be included in the calculation of the oscillation probability and can have significant effects on the oscillation probability and the determination of the neutrino parameters for all of the LBL experiments.  This is especially true for DUNE that will use the matter effects to determine the atmospheric mass ordering but it is also true for all the experiments for the determination of the size of CP violation, \cite{Mena:2004sa}. Although the density of the earth's crust varies somewhat over the baseline of all the above  LBL experiments it has been shown in \cite{Kelly:2018kmb, King:2020ydu} that a constant density approximation gives accurate enough oscillation probabilities given the statistics of these experiments. For a review of CP violation in neutrino oscillations see \cite{Nunokawa:2007qh}. \\

The neutrino oscillation probabilities for three neutrino flavors in constant medium have been exactly calculated\footnote{Use of efficient numerical packages, while excellent for generating figures and doing analysis, they do not provide detailed physics insight or understanding.}  by many authors:
Barger et al \cite{Barger:1980tf}, Zaglauer \& Schwarzer \cite{Zaglauer:1988gz},  Harrison \& Scott \cite{Harrison:1999df},  Ohlsson \& Snellman \cite{Ohlsson:1999xb, Ohlsson:1999md}, Kimura et al \cite{Kimura:2002hb, Kimura:2002wd}  as well as many approximation schemes reviewed in \cite{Barenboim:2019pfp}.
All of the exact methods first determine the squared masses of the neutrinos in the medium by solving the characteristic equation, $\text{\bf Det}(\lambda I -H)=0$, where $\lambda$'s are the eigenvalues of the Hamiltonian, $H$, and $I$ is the identity matrix. Where these exact methods differ, is how one then determine the eigenvectors associated with the eigenvalues. Using the methods implored in the above papers the  calculations of the eigenvectors are far from transparent.  The eigenvectors make up the elements of the mixing matrix for the neutrino states in the medium and with these one can easily calculate the oscillation probabilities.   In this paper, we present a different, efficient and simple way to determine the eigenvectors which is known in the mathematics community but so far has not appeared in the neutrino physics literature.   We call this method the ``adjugate method'' as it requires the calculation of the adjugate, the transpose of the co-factor matrix, of the Hamiltonian, $\text{\bf Adj}(H)$.  Because of the form of the neutrino Hamiltonian in the flavor basis  this adjugate method is particularly suited to neutrino oscillation calculations as we will show in Sec.\,\ref{sec:DiagH}. We use this method to identify a number of quantities that are invariant, that is, independent of the matter potential, in the various scenarios addressed. For the standard three flavor case these identities are known whereas for the other scenarios these identities are new such as the generalization of the Naumov-Harrison-Scott identity for four or more flavors of neutrinos.\\

The outline of this paper is as follows:  in Sec.\,\ref{sec:oscprobs} we revisit what is required to calculate the oscillation probabilities in matter with particular attention to the intrinsic CP violating term.  
Sec.\,\ref{sec:DiagH} we discuss the various methods for diagonalizing a Hamiltonian and present the ``adjugate method'' and why it is particularly suited to neutrino oscillation calculations. Sec.\,\ref{sec:3f}
is revisiting the well known three flavor case to demonstrate how simple and straight forward this method is.  Sec.\,\ref{sec:4f} is application of this method to four neutrino flavors, three active and one sterile, with particularly attention to the matter potential invariants such as the generalization of various identities, including the Naumov-Harrison-Scott identity. Sec.\,\ref{sec:3f-nsi} addresses the case of three flavors plus non-standard interactions (NSI).
Next is the conclusions, Sec.\,\ref{sec:concl}, followed by a number of very useful appendices.

\section{Calculation of the Oscillation Probabilities}
\label{sec:oscprobs}

For neutrinos propagating through a constant medium, the oscillation probabilities can be easily calculated if you know the eigenvalues, $\lambda_i$, of the Hamiltonian in the flavor bases and $ V_{\alpha i} V^*_{\beta i}$
of the normalized eigenvectors, $ V_{\alpha i} $,  no matter how complicated the Hamiltonian. The $\lambda_i$'s and  the $ V_{\alpha i} $'s are the medium equivalents of $m^2_i$ and elements of the PMNS matrix $U_{\alpha i}$ in vacuum. In general the Hamiltonian is not the same for neutrinos and anti-neutrinos, so the  $\lambda_i$ and $ V_{\alpha i} V^*_{\beta i}$ need to be calculated for both.
   With these qualities the neutrino oscillation probabilities are given by
 \begin{eqnarray}
P(\nu_\alpha \to  \nu_\beta)  & =  &
\left|~\sum_j V^*_{\alpha j} ~V_{\beta j} 
\text{e}^{-i\lambda_j L/(2E)}~\right|^2\nonumber \\
& = & \delta_{\alpha \beta} 
-4 \sum_{i>j} {\Re}( V_{\alpha i} V^*_{\beta i} \,  V^*_{\alpha j}  V_{\beta j} )
\sin^2\left( \Delta_{ij} \right) \nonumber \\
& & \quad \quad  
 -8  \sum_{i>j} {\Im}( V_{\alpha i} V^*_{\beta i} \,  V^*_{\alpha j}  V_{\beta j}) \, \sin  \Delta_{ij} \sin  \Delta_{ik} \sin \Delta_{jk} \, ,
\label{eq:prob}
\end{eqnarray}
with  $\Delta_{ij} \equiv (\lambda_i -\lambda_j)L/(4E)$. There is no sum over k and it is fixed for the calculation, typically k=1 is chosen.
The last term contains the intrinsic CP violating piece and is only non-zero when $\alpha \neq \beta$. The usual way of writing this term, as in the PDG,
$$2\sum_{i>j} 
{\Im}( V_{\alpha i} V^*_{\beta i}   V^*_{\alpha j}  V_{\beta j} )
\sin \left( 2\Delta_{ij}\right) $$
appears to have a linear dependence in $(L/E)$, but this is illusionary, as this term must be of order $(L/E)^3$ in the small (L/E) limit as can be seen directly from the first line of this expression there can be no terms linear in $(L/E)$.  Therefore, it is more informative to rewrite the CP violating term as in eq. \ref{eq:prob}, since
\begin{align}
2\sum_{i>j} 
{\Im}( V_{\alpha i} V^*_{\beta i}   V^*_{\alpha j}  V_{\beta j} )
\sin \left( 2\Delta_{ij}\right) = 8  \sum_{i>j} {\Im}( V_{\alpha i} V^*_{\beta i} \,  V^*_{\alpha j}  V_{\beta j}) \, \sin  \Delta_{ij} \sin  \Delta_{ik} \sin \Delta_{jk}
 \label{eq:CPV}
\end{align}
for any fixed k. So for n-flavors there are $n$ ways to chose k, they are all equivalent. Typically k=1 is chosen. There are only (n-1)(n-2)/2 non-zero terms on the RHS, as the terms when i=k or j=k are zero. Therefore the number of such terms is the same as the number of Dirac type phases in the n-flavor PMNS matrix.  A derivation of eq.  \ref{eq:CPV} is given in appendix \ref{sec:CPVterm}.
 Details for the case when n=4 are presented in Section \ref{sec:4f}. 

In the next section we will show how to calculate both $\lambda_i$ and $ V_{\alpha i} V^*_{\beta i}$ in a systematic, straightforward and simple way, independent of the number of neutrino flavors and independent of the complexity of the Hamiltonian. Although the method for calculating $\lambda_i$ is well known, the closely associated method for  calculating $ V_{\alpha i} V^*_{\beta i}$ has not appeared in the neutrino physics literature before and to our surprise is not well known in the particle physics community. 
Once the  $\lambda_i$'s are known, there is a simple matrix expression using the adjugate of a matrix that relates the  $\lambda_i$'s  and the Hamiltonian in flavor basis to the $ V_{\alpha i} V^*_{\beta i}$'s.  \\


\section{Diagonalizing a Hermitian Hamiltonian}
\label{sec:DiagH}

Let H be an n\,$\times$\,n Hermitian matrix that we want to diagonalize as follows:
\begin{align}
H=V \Lambda V^\dagger
\label{eq-Hdiag}
\end{align}
where $\Lambda=\text{Diag}(\lambda_1, \lambda_2, \cdots, \lambda_n)$ and $V_{\alpha i}$ is a unitary matrix such that i-th column of $V$ is a {\it normalized} eigenvector of H with eigenvalue $\lambda_i$, since $HV=V\Lambda$.

It is well known that the eigenvalues are obtained be solving the characteristic equation
\begin{align}
\text{Det}[\lambda_i I-H] =0 \,.
\label{eq-Det}
\end{align}
For $n \leq 4$ this can be done analytically, but for $n$ $>$ 4 this cannot be done analytically in general.

It is well known in the mathematics community, but rarely known in the particle physics community,  that the components of the normalized,  i$^{\text{th}}$  eigenvector are easily calculated using
\begin{align}
V_{\alpha i}V^*_{\beta i} =\frac{\text{\bf Adj}[\lambda_i I-H]_{\alpha \beta}}{\Pi_{k\neq i} (\lambda_i-\lambda_k)} \,,
\label{eq-Adj}
\end{align}
where Adj is the adjugate of the matrix\footnote{See wikipedia \href{https://en.wikipedia.org/wiki/Adjugate_matrix}{Adjugate Matrix}. The adjugate of a matrix (the transpose of the co-factor matrix) is well defined independent of whether or not the matrix is invertible.  For invertible matrices, the inverse of the matrix can be calculated using
$H^{-1} =\text{\bf Adj}[H] /\text{\bf Det}[H]$.  The matrix $(\lambda_i I-H)$ is not invertible since $\text{\bf Det}[\lambda_i I-H]=0$. }, see \cite{Denton:2019pka}, i.e. the transpose of the co-factor matrix and references therein\footnote{Note, adding $xI$ to the Hamiltonian leaves the RHS of this equation unchanged, as x is added to all eigenvalues. The physics here is that, neutrino oscillations cannot determine the absolute neutrino mass scale.}. For neutrino oscillations one needs exactly the quantities  $ V_{\alpha i} V^*_{\beta i}$ which are given directly and simply from eq.\ref{eq-Adj}.  If needed, this equation can also be used to calculate the eigenvectors in a simple, straight forward manner, up to an overall phase which is arbitrary. The quantity $ V_{\alpha i} V^*_{\beta i}$ given by eq. \ref{eq-Adj} is independent of this arbitrary choice but can be used to calculate the magnitude, 
$ | V_{\alpha i} |$,  and relative phase of each component of an eigenvector, since this phase is given by  $ V_{\alpha i} V^*_{\beta i}/(| V_{\alpha i} | | V_{\beta i}|) $.
 A physicist's style sketch of why  eq.\ref{eq-Adj} is correct is given in appendix \ref{sec:adjproof}.

With $\lambda_i$'s and the $V_{\alpha i}V^*_{\beta i} $'s  constructed this way, eq.\,\ref{eq-Hdiag}
is satisfied and 
\footnote{While the first of these is easily checked from eq. \ref{eq-Adj}, the latter is most easily checked in numerical calculations using $ \sum_\beta V_{\alpha i}V^*_{\beta i} V_{\beta j}V^*_{\gamma j }  = \delta_{ij} V_{\alpha i}V^*_{\gamma i} $. }
\begin{align}
 \sum_i V_{\alpha i}V^*_{\beta i}  = \delta_{\alpha \beta}   \quad \text{and} \quad   \sum_\alpha V_{\alpha i}V^*_{\alpha j} = \delta_{ij}\,.
 \end{align}
 Therefore the eigenvectors constructed this way are normalized and orthogonal.\\

Even less well known until recently, even in the mathematics community, is that the magnitudes of the elements of V can be simple, obtained using the eigenvector-eigenvalue identity: \cite{Denton:2019ovn} \& \cite{Denton:2019pka},
\begin{align}
|V_{\alpha i}|^2 =\frac{\Pi_{l} (\lambda_i-\lambda_l(h_\alpha ))}{\Pi_{k\neq i} (\lambda_i-\lambda_k)}
\label{eq-Esq}
\end{align}
where $h_\alpha$ is the principal minor of H with the $\alpha$-th row and column deleted.  $\lambda_l(h_\alpha)$ are the eigenvalues of $h_\alpha$, i.e. the solutions to 
$ \text{Det}[\lambda_lI-h_\alpha]=0$. Constructed this way $\sum_i |V_{\alpha i}|^2 =  1= \sum_\alpha |V_{\alpha i}|^2$, i.e. they are normalized appropriately.  Eq. \ref{eq-Esq} is mathematically equivalent to eq. \ref{eq-Adj} when $\alpha=\beta$, however the square of the diagonal elements are more simply obtained using eq. \ref{eq-Esq}. For standard three flavor oscillations in matter, eq.\ref{eq-Esq} is sufficient, see  \cite{Denton:2019pka}, however for more complicated scenarios like 3+1 or NSIs then
 eq. \ref{eq-Adj}  is required.  Eq. \ref{eq-Adj} gives the most straight forward and systematic way to evaluate the magnitude and relative phase of all components of the eigenvectors and gives directly the quantities,$ V_{\alpha i} V^*_{\beta i}$, for neutrino oscillation calculations. \\

Therefore, for the evaluation of the eigenvalues we need, $\text{\bf Det}(\lambda I-H)$, and for the eigenvectors, $\text{\bf Adj}(\lambda I-H)$ is also needed, see eq. \ref{eq-Det} \& eq.\ref{eq-Adj}.  We first note that both are polynomials in $\lambda$ of degree $n$ and (n-1) respectively, i.e.
\begin{align}
\text{\bf Det}[\lambda I-H] &\equiv \lambda^{n} +\lambda^{n-1} d_1+\cdots+d_n
\label{eq:dis} \\
\text{\bf Adj}[\lambda I-H] &\equiv \lambda^{n-1} A_1+\lambda^{n-2} A_2+\cdots+A_n \,.
\label{eq:Ais}
\end{align}
where the $d_i$'s are scalars and the $A_i$'s are  n\,$\times$\,n Hermitian  matrices.

There are at least five of different algebraic methods that can be used to calculate  $\text{\bf Det}[\lambda I-H]$ and $\text{\bf Adj}[\lambda I-H]$, we review them here:
\begin{enumerate}
\item  Using the {\bf Leibniz 
formulae}\footnote{ See wikipedia  \href{https://en.wikipedia.org/wiki/Leibniz_formula_for_determinants 
}{Leibniz for n\,$\times$ \,n}} 
 to calculate the determinant and elements of the adjugate of $(\lambda I-H)$. However this standard method does not, in general, give the simplest results without significant additional manipulations.
 
\item {\bf Le\,Verrier-Faddeev algorithm:} a more general method  is the very simple and recursive  Le\,Verrier-Faddeev algorithm\footnote{See wikipedia \href{https://en.wikipedia.org/wiki/Faddeev–LeVerrier_algorithm}{Faddeev–LeVerrier algorithm}.  }, see also \cite{Hou:1998jdv},  given by 
\begin{eqnarray}
&A_1 \equiv I_n  \quad  \quad &d_1=-\text{Tr}(H A_1) \notag \\
&  &\, \quad = -\text{Tr}(H)=-\sum_j \lambda_j \notag \\
&A_2= HA_1+d_1I_n. \quad & \quad d_2=-(1/2) \text{Tr}(H A_2)  \notag \\
&~\quad = H-\text{Tr}(H)I_n \quad & ~\quad =(1/2)( \text{Tr}^2(H)-\text{Tr}(H^2))=\sum_{j1>j2} \lambda_{j1} \lambda_{j2} \notag \\ [-4mm]
&\vdots \quad & \quad \quad  \vdots  \notag \\
&A_n=HA_{n-1} +d_{n-1} I_n \quad & \quad d_n=-(1/n) \text{Tr}(H A_n)\notag \\
&\, =(-1)^{n-1}\text{\bf Adj}[H] & \, \quad \quad =(-1)^n \text{\bf Det}(H)=(-1)^n\Pi^n_{j=1} \lambda_j  \,.
\end{eqnarray}
The next iteration gives 
$$A_{n+1}=(-1)^{n-1}[H\text{\bf Adj} (H)- \text{\bf Det}(H)I_n]=0,$$ 
which is related to the Caley-Hamiltonian identity. Then $d_{n+1}=0$ follows, terminating the series. As expected, from eq. \ref{eq:dis}, the d's are the combination of traces of $H^m$ (m=1,\dots, n) such that $d_m =  (-1)^m \sum_{ j1>...>jm } \lambda_{j1} \cdots \lambda_{jm}$. In appendix \ref{app:LeVF} the d's and A's are explicitly given for  n$\leq$6.  

It is worth noting here the following relationship between the trace of the A's and the d's:
$\text{\bf Tr}[A_{m+1}] =(n-m)d_m$,
valid for m=0, $\cdots$, n. (using $d_0 \equiv 1$), and therefore giving
\begin{align}
\text{\bf Tr}[ \text{\bf Adj}[\lambda I -H] ]=\frac{\partial}{\partial \lambda} \, {\bf Det}[\lambda I -H]
\end{align}
for any $\lambda$.

\item
{\bf Lagrange method:}  the  results of the Le\,Verrier-Faddeev algorithm can be re-arranged, see appendix \ref{app:LeVF2}, as 
\begin{align}
\text{\bf Adj}(\lambda_i I-H) 
&= 
{\bf \Pi}_{j\neq i} (H-\lambda_j I)  \,.
\label{eq:Adj.eq.Pi}
\end{align}
 This is the Lagrange method that has been used in many papers: \cite{Harrison:2002ee}, \cite{Fong:2022oim}.
The LHS of this equation only depends on the eigenvalue $\lambda_i$ whereas the RHS depends on all the other eigenvalues, $\lambda_j$'s with $ j \neq i$.  The LHS and RHS of eq. \ref{eq:Adj.eq.Pi} are the two extremes of the many ways to write this expression given the many relationships between the traces of the powers of the Hamiltonian and products of the eigenvalues, e.g. $\sum_i  \lambda_i=\text{tr}(H)$, 
$\sum_{i>j}  \lambda_i  \lambda_j =\frac1{2}[\text{tr}^2(H)- \text{tr}(H^2)]$, etc.
This implies that the denominator of eq. \ref{eq-Adj} can also be written as
$
\Pi_{j \neq i} (\lambda_i-\lambda_j) = \text{\bf Tr}[ {\bf \Pi}_{j\neq i} (H-\lambda_j I) ] 
$.  
\\

For numerical calculations, the  Le\,Verrier-Faddeev algorithm and using the Lagrange expression
require a similar numbers of complex matrix operations. However, the  Le\,Verrier-Faddeev algorithm gives coefficients of $\lambda^k$ for both the determinant and the adjugate of  $(\lambda I -H)$ in one very simple iterative procedure. For purely numerical calculations, especially if the matrices are large, there exist many very efficient numerical packages.  However for analytic neutrino oscillation calculations where the number of neutrino flavors is $\leq$6, we will now argue that using the  adjugate of $(\lambda I -H)$ is simpler, due to the form of the Hamiltonian, as we will see in the next item.

\item 
{\bf Adjugate method:} for neutrino oscillation calculations one can use the special form of the Hamiltonian in the flavor basis as follows:
\begin{align}
H &=H_0  + H_1
\end{align}
where $H_0$ is the vacuum Hamiltonian in the flavor basis, so that, 
\begin{align}
H_0 &=U X U^\dagger \notag \\
\text{with} \quad X &=\text{\bf diag}(x_1, x_2, \cdots, x_n)
\end{align}
where $x_i=m^2_i/(2E)$. The adjugate of $H_0$ is simply given by
\begin{align}
\text{\bf Adj}[H_0]_{\alpha \beta} &= \sum_i U_{\alpha i} U^*_{\beta i} \,  \Pi_{j\neq i} x_j \,,
\label{eq:AdjH0}
\end{align}
which follows from $\text{\bf Adj}[U X U^\dagger] =U\text{\bf Adj}[X] U^\dagger$.
 $H_1$ contains the matter potential which for most cases is a sparse matrix in the flavor basis.  For the standard 3 flavor case only one diagonal element of $H_1$ is non-zero and for 3+1 flavors only 2 diagonal elements are non-zero.  If $H_1$ is sparse, then calculation of the  $\text{\bf Adj}[H]$ is simple and straight forward.  A similar argument applies to  $\text{\bf Det}[H]$.\\

However, what is actually needed is $\text{\bf Det}[\lambda I-H]$ and $\text{\bf Adj}[\lambda I-H]$, but these can be trivially calculated from $\text{\bf Det}[H]$ and $\text{\bf Adj}[H]$, since
\begin{align}
\lambda I-H= U(\lambda I-X) U^\dagger -  H_1 \, .
\end{align}
Therefore, replacing  $x_i \rightarrow (x_i -\lambda)$  in  $\text{\bf Det}[H]$ and $\text{\bf Adj}[H]$ one obtains the required results,
since
\begin{align}
\text{\bf Det}[\lambda I-H(x_i)] &=(-1)^n ~ \text{\bf Det}[H(x_i-\lambda)] \,, \notag \\[3mm]
\text{\bf Adj}[\lambda I- H(x_i)] &=(-1)^{n-1} ~ \text{\bf Adj}[H(x_i-\lambda)] \,.
\end{align}
Therefore one only needs to calculate $\text{\bf Det}[H]$ and $\text{\bf Adj}[H]$, a considerable simplification.
We call the method given here, the ``adjugate method,'' and it is the most straight forward and systematic for analytic neutrino oscillation calculations. This happens because the full Hamiltonian can be split into an easily diagonalized part, $H_0$, and an additional piece, $H_1=H-H_0$, that is a sparse matrix.

\end{enumerate}

All of these methods, as well as combinations, can be used to find the required results, however, the last method given above, adjugate method, requires the least algebraic computation and gives simple expressions for neutrino oscillations because of the form of the Hamiltonian. This is the method that will be used in the rest of this paper for the numerator of eq. \ref{eq-Adj}.\\


A comment about the denominator of adjugate equation, eq. \ref{eq-Adj}, is in order as well.  To evaluate this denominator it appears that you need all the eigenvalues of H, not just the eigenvalue of the eigenvector one is calculating. This is illusionary, as one can easily show that
\begin{align}
{\bf \Pi}_{j\neq i} (\lambda_i-\lambda_j )  &=  \frac{\partial}{\partial \lambda} \, {\bf Det}[\lambda I -H]\,
\biggr|_{\lambda=\lambda_i} 
=  n\lambda^{n-1}_i +(n-1)\lambda^{n-2}_i d_1+\cdots+d_{n-1}  \,,
\label{eq:Denom} 
\end{align}
 see  for example ref.  \cite{Denton:2019pka}.  ${\bf Det}[\lambda I -H]$ is just the characteristic polynomial and the coefficients are given by the $d_i$'s of eq. \ref{eq:dis}.
Thus, if you know the position of a zero of the characteristic polynomial and the slope of this polynomial at that zero, one has enough information using eq. \ref{eq-Adj}  to determine that eigenvector i.e. in physicist's language the determination of the eigenvector is local in $\lambda$, one does not need to know all the other eigenvalues. That is, the adjugate equation, eq. \ref{eq-Adj}, could have been written as
\begin{align}
V_{\alpha i}V^*_{\beta i} = \biggr( \text{\bf Adj}[\lambda I-H]_{\alpha \beta}\biggr/  \frac{\partial}{\partial \lambda} \, {\bf Det}[\lambda I -H] \biggr) \,  \biggr|_{\lambda=\lambda_i}  \,,
\label{eq-Adj2}
\end{align}
which manifestly only depends on the single eigenvalue of $H$,  $\lambda_i$, associated with the eigenvector given by $V_{\alpha i}V^*_{\beta i}$.
Of course frequently one may need all eigenvalues, if so, then  RHS of eq. \ref{eq:Denom}  maybe the simplest route but it is not necessary for the determination of a single eigenvector.  In addition eq. \ref{eq:Denom} can also be used as a cross check.

Alternatively, since we require $\sum_{\alpha} |V_{\alpha i}|^2=1$, then the denominator of eq. \ref{eq-Adj} is clearly equal to the  $\text{\bf Tr}[ \text{\bf Adj}[\lambda_i I-H]]$, see appendix \ref{sec:adjproof},
as all of these are equivalent since
$${\bf \Pi}_{j\neq i} (\lambda_i-\lambda_j )  =\frac{\partial}{\partial \lambda} \, {\bf Det}[\lambda I -H]\,
\biggr|_{\lambda=\lambda_i}  = \text{\bf Tr}[ \text{\bf Adj}[\lambda_i I-H]] 
=\text{\bf Tr}[ {\bf \Pi}_{j\neq i} (H-\lambda_j I)] .$$

\subsection{Relationship to previous works}

The method used in the papers of Kimura et al. (KTY), \cite{Kimura:2002hb} \& \cite{Kimura:2002wd}, specifically for three neutrinos, is closely related but not exactly the calculation of $\text{\bf Adj}(\lambda_i I-H)$ for a 3x3 matrix. In the  KTY papers the adjugate equation, eq. \ref{eq-Adj}, never appears, nor do the words ``adjugate'' or ``classical adjoint'' or "adjunct" appear, even though some of the results of their papers are closely related to what we find for three neutrinos in the next section.   In fact, the results given in the KTY papers are midway\footnote{It is easy to show that the numerator of eq. 39 of Kimura et al \cite{Kimura:2002wd}, the key equation of that paper, can be rewritten with the eigenvalues $(\lambda_i, \lambda_j, \lambda_k)$ with (i,j,k) all different, as 
$$-\lambda_i(\lambda_j+\lambda_k)I_3 + \lambda_i H+H^{-1}\text{\bf Det}(H) =\lambda_j\lambda_k I_3  - (\lambda_j+\lambda_k) H+H^2 =  \lambda^2_i  I_3+ \lambda_i (H-\text{\bf tr(}H) I_3)+\text{\bf Adj}(H). $$
The left expression from KTY,  depends explicitly on all 3 eigenvalues, $(\lambda_i, \lambda_j, \lambda_k)$, the middle, which comes from $ {\bf \Pi}_{j\neq i} (H-\lambda_j I)$, \cite{Harrison:2002ee}, depends explicitly on 2 eigenvalues,  $(\lambda_j, \lambda_k)$, whereas the right, which comes from $\text{\bf Adj}(\lambda_i I-H)$, depends explicitly only on {\it the} eigenvalue, $\lambda_i$.   Here, (2E) is set to 1 for simplicity. } between the calculations using
$\text{\bf Adj}(\lambda_i I-H)$ given in the next section and $ {\bf \Pi}_{j\neq i} (H-\lambda_j I)$ of
ref. \cite{Harrison:2002ee}.  There are multiple paths to extending the KTY method to more than three neutrinos, whereas  $\text{\bf Adj}(\lambda_i I-H)$  and $ {\bf \Pi}_{j\neq i} (H-\lambda_j I)$  are uniquely determined.    The methods based on the Caley-Hamilton Theorem,
$ {\bf \Pi}^n_{j=1} (H-\lambda_j I)=0$, such as papers \cite{Ohlsson:1999xb}, \cite{Ohlsson:1999md} \& \cite{Bustamante:2019ggq},  are more distantly related and are more complicated to implement. 
\\

Our method is very simple, general and systematic, and can be applied to more than 3 neutrinos or to more complicated matter potentials than the method in KTY.  In some sense it can be considered a modification of the above methods that allows a straight forward generalization to more than 3 neutrinos or more complicated matter potentials.

\newpage

\section{3 Flavors in Matter: revisited}
\label{sec:3f}

In this section we have used the adjugate method to calculate of the eigenvectors, i.e $V_{\alpha i}V^*_{\beta i}$, and to reproduces the known results for 3 flavors. This not only demonstrates the simplicity of this method but also by reproducing known results confirms its validity for this important and non-trivial example.

 For three flavors in matter the Hamiltonian is simply 
\begin{align}
(2E) H_{3x3} &=U M^2 U^\dagger +  \text{Diag}(a,0,0) \,,
\label{eq:3fH}
\end{align}
where $U$ is the PMNS matrix, $M^2$ the mass matrix squared, $M^2=\text{Diag}(m^2_1,m^2_2,m^2_3)$ and $a=2\sqrt{2} G_F N_e E$  is the Wolfenstein matter potential, \cite{Wolfenstein:1977ue}, where  $E$ is the energy of the neutrino in the rest frame of the matter. The term proportional to the number density of neutrons, has been subtract from all diagonal elements of the Hamiltonian, eq. \ref{eq:3fH}, as terms proportional to the identity matrix in the Hamiltonian do not effect oscillations. Now
\begin{align}
\text{Det}[(2E) H_{3x3}] &=\Pi_i\, m^2_i+a  \sum_{i} |U_{ei}|^2 \,  \Pi_{j\neq i}m^2_j \, .
\label{eq:detH3x3}
\end{align}
Useful identities for this calculation can be found in appendix \ref{sec:SUI}. Therefore by replacing $m^2_i \rightarrow (m^2_i-\lambda)$ we obtain the characteristic equation\footnote{This is equivalent to $$ \lambda^3-\lambda^2 Tr[(2E)H]  +\lambda (Tr^2[(2E)H]-Tr[(2E)^2H^2])/2 -\text{Det}[(2E)H] =0.$$ }
\begin{align}
\text{\bf Det}[\lambda I-(2E) H_{3x3}]&=
 \lambda^3-\lambda^2 \biggr( \sum_i m^2_i +a \biggr) 
+\lambda \biggr( \sum_{i>j} m^2_i m^2_j +a \sum_i m^2_i (1-|U_{ei}|^2) \biggr) \notag  \\
 &  \quad -\biggr(\Pi_i m^2_i+ a  \sum_{i} |U_{ei}|^2 \,  \Pi_{j\neq i}m^2_j  \biggr) = 0 \, .
 \label{eq-det3}
\end{align}
Note that if one sets $(m^2_1,m^2_2,m^2_3)=(0, \Delta m^2_{21}, \Delta m^2_{31})$, eq. \ref{eq-det3} reduces to the results of Zaglauer-Schwarzer, \cite{Zaglauer:1988gz}, and there are significant simplifications as any term with $m^2_1$ is zero. In this paper we prefer not to do this so as to maintain the symmetry between the $m^2_i$. The exact solutions to this cubic equation are given in appendix  \ref{sec:cubic}. \\

To obtain the eigenvectors, we need both the numerator and the denominator of  the adjugate eqn. \ref{eq-Adj}.
The denominator can be simply evaluated using $\text{\bf Det}^\prime[\lambda I-(2E) H_{3x3}]$, see eqn. \ref{eq:Denom}, with $\lambda$ being the eigenvalue of the eigenvector.  Note it is linear in the matter potential if one uses only the eigenvalue of the eigenvector being calculated.

The numerator is easily evaluated from
$\text{\bf Adj}[(2E) H_{3x3}$, which is super simple to calculate and given by
\begin{align}
\text{\bf Adj}[(2E) H_{3x3}]_{\alpha \beta} &=  \sum_{i}  U_{\alpha i}U^*_{\beta i}  \,  \Pi_{j\neq i} \, m^2_j   \,
+a \left\{\begin{array}{ll}
0& \alpha=e ~~ or ~~ \beta=e \\
 \sum_i m^2_i |U_{\tau i}|^2 & \alpha=\beta = \mu \\
 \sum_i m^2_i |U_{\mu i}|^2 & \alpha= \beta = \tau \\
- \sum_i m^2_i U_{\alpha i}  U^*_{\beta i} \quad &  \alpha \neq \beta, \alpha \neq e, \beta \neq e,  
 \end{array}
 \right.  \, .
\label{eq:adjH3x3} 
\end{align}
Again each sum has three terms (two if $m^2_1=0$).  The matter potential independent part here is most easily calculated using the identity given in eqn. \ref{eq:AdjH0} whereas the matter dependent part is trivial. Note, $\text{\bf Adj}[(2E) H_{3x3}]$ is also linear in the matter potential. 

 The eigenvectors are then calculated as follows:
\begin{align}
V_{\alpha i}V^*_{\beta i} & =\frac{\text{\bf Adj}[\lambda_i I-(2E) H_{3x3}]_{\alpha \beta}}{\Pi_{l\neq i} (\lambda_i-\lambda_l)} 
\equiv \frac{\lambda^2_i \delta_{\alpha \beta} - \lambda_i S_{\alpha \beta} + T_{\alpha \beta}}{\Pi_{l\neq i} (\lambda_i-\lambda_l)}  \, ,
\label{eq-Adj3}
\end{align}
where $\text{\bf Adj}[\lambda I-(2E) H_{3x3}]$ can be easily obtained from $\text{\bf Adj}[(2E) H_{3x3}]$ by replacing $m^2_i \rightarrow (m^2_i-\lambda)$.
Thus $S_{\alpha \beta}$ and $T_{\alpha \beta}$ are simply,
\begin{align}
S_{\alpha \beta}& =  \sum_i m^2_i (\delta_{\alpha \beta} - U_{\alpha i}U^*_{\beta i} ) +a \delta_{\alpha \beta}(1-\delta_{\alpha e}) \quad \text{and} \quad T_{\alpha \beta} =\text{Adj}[(2E)H_{3x3}]_{\alpha \beta} \, .
\end{align}
 
 The diagonal terms   ($\beta = \alpha $) are then easily obtained as 
 \begin{align}
 | V_{\alpha i} |^2 &= \frac{ \lambda^2_i - \lambda_i S_{\alpha \alpha} + T_{\alpha \alpha} }{\Pi_{l\neq i} (\lambda_i-\lambda_l)} 
  \quad \text{with} \quad  S_{\alpha \alpha} =  \sum_i m^2_i (1 - |U_{\alpha i}|^2 ) +a (1-\delta_{\alpha e})
  \notag \\ 
 \quad \text{and} \quad T_{\alpha \alpha} &= 
 \sum_{i}  U_{\alpha i}U^*_{\beta i}  \,  \Pi_{j\neq i}m^2_j   
 +a \left\{\begin{array}{cl}
0& \alpha=e  \\
 \sum_i m^2_i |U_{\tau i}|^2 & \alpha=\mu \\
 \sum_i m^2_i |U_{\mu i}|^2 & \alpha=\tau \\
 \end{array}
 \right. 
 \end{align}
 
 In the adjugate method where we only use the eigenvalue of the eigenvector we are calculating, then both the numerator and denominator are quadratic polynomials in the eigenvalue with coefficients that are linear in the matter potential. Other formulations such as the Lagrange method do not have such a simple dependence, but of course, they give the same numerical values.
 
 \subsection{Matter potential invariant quantities for three flavors}
 
The elements of the Hamiltonian given by eq. \ref{eq:3fH} are nearly all independent of the matter potential, except for the $H_{ee}$ component. In this section we will explain what the physics implications of this independence has  on the oscillation probabilities.  First we note that because we can re-phase the flavor states we must also take this into account when identify these physics implications.

First, the diagonal elements of the Hamiltonian give the ``mass'' of the flavor states:
\begin{align}
m^2_\alpha = \sum_i |U_{\alpha i}|^2 m^2_i + a \delta _{\alpha e}= \sum_i |V_{\alpha i}|^2 \lambda_i \,.
\end{align} 
 Note the ``mass'' of $\nu_e$ is the exception here and depends linearly on the  matter potential whereas the "masses" of $\nu_\mu$ and $\nu_\tau$ are independent of the matter potential.
 
 Second the square of the off-diagonal elements of the Hamiltonian ($|H_{\alpha \beta}|^2$, $\alpha \neq \beta$)  are related to the appearance oscillation probabilities in the small L/E limit. Therefore in this limit the coefficients of the $(L/E)^2$ terms in the oscillation probabilities, i.e.  CP or T conserving parts, must be invariant for $\alpha\neq \beta$, \cite{Harrison:2002ee}:
\begin{align}
  \sum_{j>k} {\cal R}(V_{\alpha j}V^*_{\beta j} V^*_{\alpha k}V_{\beta k} )~ (\Delta \lambda_{jk})^2 
 & =  \sum_{j>k} {\cal R}(U_{\alpha j}U^*_{\beta j} U^*_{\alpha k}U_{\beta k} )~ (\Delta m^2_{jk})^2\,.
 \label{eq:3CPC}
 \end{align}
This identity identity is easily checked using the expressions given in the previous subsection and can be used as cross check.

Another identity, which is not independent of the above, is associated with the disappearance probabilities:
\begin{align}
 \sum_{j>k} |V_{\alpha j}|^2 |V_{\alpha k}|^2 (\Delta \lambda_{jk})^2 
 & = \sum_{j>k} |U_{\alpha j}|^2 |U_{\alpha k}|^2 (\Delta m^2_{jk})^2
 \label{eq:3disapp}
 \end{align}
 This identity guarantees that in $L/E \rightarrow 0$ limit, the disappearance oscillation probabilities is independent of matter effects i.e. the the coefficients of the  terms of order  $(L/E )^2$ are equal  in matter and vacuum. This identity follows from the matter potential independence of $\sum_{\beta \neq \alpha}| H_{\alpha \beta}|^2$.\\

What about the  intrinsic CP violating (CPV) or T violating part of the appearance probabilities?  The CP or T violating part of the oscillation probabilities appears first an $(L/E)^3$ term given by
\begin{align}
\Im (V_{\alpha i}V^*_{\beta i} V^*_{\alpha j}V_{\beta j} ) \Delta \lambda_{ij} \Delta \lambda_{ik}
 \Delta \lambda_{jk} 
  \label{eq:nhs}
 \end{align}
and is also invariant. This is the  Naumov-Harrison-Scott (NHS) invariant, \cite{Naumov:1991ju} \& \cite{Harrison:1999df} which follows from the invariance of $\Im( H_{e \mu} H_{\mu \tau} H_{\tau e})$.
Note, rephasing of the flavor states does not change this invariant. $\Re( H_{e \mu} H_{\mu \tau} H_{\tau e})$ gives no new invariant.  The NHS identity can be easily derived from our expressions as follows:
for $\alpha \neq \beta$ and $i \neq j$, we obtain
\begin{align}
 V_{\alpha i}V^*_{\beta i} V^*_{\alpha j}V_{\beta j} & =  \frac{ \lambda_i \lambda_j |S_{\alpha \beta}|^2 -\lambda_i S_{\alpha \beta}  T^*_{\alpha \beta}  -\lambda_j  S^*_{\alpha \beta}  T_{\alpha \beta} 
 +|T_{\alpha \beta}|^2  }{\Pi_{l\neq i} (\lambda_i-\lambda_l)~ \Pi_{k\neq j} (\lambda_j-\lambda_k)}  
 \end{align}
and therefore  the Jarlskog invariant, \cite{Jarlskog:1985ht}, is given by
\begin{align}
\Im (V_{\alpha i}V^*_{\beta i} V^*_{\alpha j}V_{\beta j} )& = \frac{(\lambda_j -\lambda_i)~ {\Im}(S_{\alpha \beta}  T^*_{\alpha \beta})}{\Pi_{l\neq i} (\lambda_i-\lambda_l)~ \Pi_{k\neq j} (\lambda_j-\lambda_k)}  
\notag \\[2mm]  &  =  \frac{ \Delta m^2_{ij} \Delta m^2_{ik} \Delta m^2_{jk}}{ \Delta \lambda_{ij} \Delta \lambda_{ik} \Delta \lambda_{jk}} ~~ {\Im}(U_{\alpha i}U^*_{\beta i} U^*_{\alpha j}U_{\beta j} ) 
\end{align}
where (i,j,k) are all different. This important identity as be used in many places to understand the effects of matter on CP and T violating oscillation probabilities, see for example \cite{Yokomakura:2000sv},  \cite{Parke:2000hu}, \cite{Nunokawa:2007qh}, \cite{Wang:2019yfp} and \cite{Denton:2019yiw}.  \\

This next identity follows from the fact that matter potential is invariant under rotations in the $\mu$ and $\tau$ sectors. It is simple algebra to show that Kimura-Takamura-Yokomakura (KTY) identity, \cite{Kimura:2002wd},
 \begin{align}
 |V_{e1}|^2   |V_{e2}|^2  |V_{e3}|^2 (\Delta \lambda_{21} \Delta \lambda_{31} \Delta \lambda_{32})^2
 \label{eq:kty}
 \end{align}
 is an invariant, i.e. independent of the matter potential.  One only has to use that the $\lambda_i$'s satisfy the 3-flavor characteristic equation, eq. \ref{eq-Det} to prove this identity.  When taking the square root of this identity care must be taken with the sign (phase) which depends on conventions, see \cite{Denton:2021vtf}. \\

The Toshev invariant, \cite{Toshev:1991ku}, is obtained by dividing the NHS identity by the square root of the KTY identity, i.e.
\begin{align}
\left(\frac{\Delta \lambda_{ij} \Delta \lambda_{ik} \Delta \lambda_{jk}}{|\Delta \lambda_{ij} \Delta \lambda_{ik} \Delta \lambda_{jk}|}\right)\frac{ \Im (V_{\alpha i} V^*_{\beta i} V^*_{\alpha j} V_{\beta j})} { |V_{e1} V_{e2} V_{e3}| } \,  ,
\label{eq:Toshev}
\end{align}
where (i,j,k) are all different. Note, the piece of this expression in $\left(\cdots \right)$ is either $\pm 1$. Some care is needed with the phases of $V_{ei}$ and signs of $\Delta \lambda_{ij}$ when taking the square root KTY identity, as noted in reference \cite{Denton:2021vtf}.  Using the PDG PMNS matrix with all $\Delta \lambda_{ij}>0$ one obtains the usual Toshev invariant (the invariance of $\sin 2 \theta_{23} \sin \delta$), however departures from the PDG conventions can introduce an additional sign. Eq. \ref{eq:Toshev} is the PMNS convention independent Toshev invariant which will be useful if a different PMNS convention than that of the PDG is used \cite{Denton:2020igp}.

\newpage

\newcommand{\bigzero}{\mbox{\normalfont\Large\bfseries 0}}
\newcommand{\rvline}{\hspace*{-\arraycolsep}\vline\hspace*{-\arraycolsep}}

\section{4 Flavors in Matter:}
\label{sec:4f}

\def\ndelta{\eta}

For four flavors in matter the Hamiltonian is simply 
\begin{align}
(2E) H_{4x4} &=U M^2 U^\dagger +  \text{Diag}(a,0,0,b) 
\end{align}
where $U$ is the 4 flavor PMNS matrix, $M^2$ the mass matrix squared, $M^2=\text{Diag}(m^2_1,m^2_2,m^2_3,m^2_4)$, ``a'' is the $N_e$ matter potential with $a=2\sqrt{2} G_F N_e E_\nu$ and 
 ``b'' is the $N_n$  matter potential with $b=\sqrt{2} G_F N_n E_\nu$ where $N_e$ and $N_n$ are the number density for electrons and neutrons respectively \footnote{ $bI$ has been added to $H$.}.

The determinant is easily calculated as
\begin{align}
 \text{Det}[(2E) \, H_{4x4}] = & \Pi_i m^2_i + a \sum_i |U_{ei}|^2 \, \Pi_{j\neq i}m^2_j + b\sum_i |U_{si}|^2 \,  \Pi_{j\neq i}m^2_j \notag 
 \\
 &+ab \sum_{i>j} |U_{e i} U_{s j}-U_{e j} U_{s i}|^2 \, \Pi_{k \neq i, \,k \neq j} m^2_k  \,.  \label{eq:det4}
\end{align}
Here we use $|U_{\mu i} U_{\tau j}-U_{\mu j} U_{\tau i}|^2 = |U_{e k} U_{s l}-U_{e l} U_{s k}|^2 $ with $(k,l)\neq (i,j)$.
Then the characteristic equation 
\begin{align}
\text{Det}[\lambda I_4-(2E)H_{4x4}] =\lambda^4-f_1 \lambda_3 +f_2 \lambda^2 -f_3 \lambda +f_4 =0
\end{align}
is easily obtained by making the $m^2_i \rightarrow (m^2_i -\lambda)$ substitution in \ref{eq:det4}, which gives
\begin{align}
 f_1 &=  
  \sum_i m^2_i +a +b \\
 f_2 &
 = \sum_{i>j} m^2_i m^2_j +a \sum_i m^2_i (1-|U_{ei}|^2)
 +b\sum_i m^2_i(1- |U_{si}|^2) +ab 
 \end{align}
 \begin{align}
 f_3 
 &= \sum_i \, \Pi_{j\neq i} m^2_j\notag \\&   +a \sum_{i>j} m^2_i m^2_j(1-|U_{ei}|^2 -|U_{ej}|^2)  +b  \sum_{i>j}m^2_i m^2_j(1-|U_{si}|^2 -|U_{sj}|^2) \notag  \\&  +ab\sum_{i} m^2_i (1-|U_{ei}|^2 -|U_{si}|^2 ) \\
 f_4&=\text{Det}[(2E)H_{4x4}] =\Pi_i m^2_i + a \sum_i |U_{ei}|^2 \, \Pi_{j\neq i}m^2_j + b\sum_i |U_{si}|^2 \,  \Pi_{j\neq i}m^2_j 
  \notag  \\
 & \hspace*{3cm} 
 +ab \sum_{i>j} |U_{e i} U_{s j}-U_{e j} U_{s i}|^2 \, \Pi_{k \neq i, \,k \neq j} m^2_k  
  \, .
 \end{align}
See appendix \ref{sec:SUI} for some useful identities.
This is in agreement with ref. \cite{Li:2018ezt},

The solution to the 4 flavor characteristic equation is as follows:
\begin{align}
 0&=\lambda^4-f_1 \lambda^3 +f_2 \lambda^2 -f_3 \lambda +f_4&\\
 &=(\lambda^2 -.\frac1{2}(f_1-\sqrt{f_1^2+4(y-f_2)}~)\lambda+ \frac1{2}(y+\sqrt{y^2-4f_4})) \\
 & \times     
 (\lambda^2-\frac1{2}(f_1+\sqrt{f_1^2+4(y-f_2)} ~)\lambda+\frac1{2}(y-\sqrt{y^2-4f_4}))
 \end{align}
where
\begin{align}
y^3 -f_2 y^2 +(f_3f_1-4f_4) y + f_4(4f_2-f^2_1)-f^2_3=0\, .
\label{eq:4fcubic}
\end{align}
The solution to this cubic equation can be found in the appendix \ref{sec:cubic}.
Therefore the solutions are then simply given by the solutions of the quadratics
\begin{align}
\lambda^2 -.\frac1{2}(f_1-\sqrt{f_1^2+4(y-f_2)}~)\lambda+ \frac1{2}(y+\sqrt{y^2-4f_4}) =0 \\   
 \lambda^2-\frac1{2}(f_1+\sqrt{f_1^2+4(y-f_2)} ~)\lambda+\frac1{2}(y-\sqrt{y^2-4f_4})=0 \,.
\end{align}
The solutions to four flavor cubic, eq. \ref{eq:4fcubic}, is the challenging part to solving the four flavor characteristic equation.

To obtain the eigenvectors, we first calculate the  $\text{\bf Adj}[(2E) H_{4x4}]$, which is given by
\begin{align}
& \text{\bf Adj}[(2E)H_{4x4}]_{\alpha \beta} = \sum_i U_{\alpha i}U^*_{\beta i} ~ \Pi_{j \neq i} m^2_j 
 \notag \\[2mm]
& \hspace*{0.5cm}
 +a\begin{pmatrix}
  \begin{matrix}
  0 
  \end{matrix}
  & \rvline & 0 ~~~~~~ 0 ~~~~~~ 0 \\
\hline
  \begin{matrix} 0 \\0\\  0 \end{matrix} & \rvline &
  \begin{matrix}
 {\large  \text{\bf Adj}[{\cal W}_{\mu, \tau, s}] }
  \end{matrix}
\end{pmatrix}_{\alpha \beta}
+b\begin{pmatrix}
  \begin{matrix}
 {\large  \text{\bf Adj}[{\cal W}_{e,\mu,\tau}] }
  \end{matrix}
   & \rvline & \begin{matrix} 0\\0\\ 0 \end{matrix} 
 \\ 
    \hline
   \begin{matrix}
  0  ~~~~~ 0 ~~~~~0 
  \end{matrix}
  & \rvline & 0
 \end{pmatrix}_{\alpha \beta}
 +ab\begin{pmatrix}
   0 & \rvline & 0 \hspace{10mm} 0 & \rvline & 0 \\
\hline
  \begin{matrix} 0\\0 \end{matrix} & \rvline &
  \begin{matrix}
 {\large  \text{\bf Adj}[{\cal W}_{\mu,\tau}] }
  \end{matrix}
  &  \rvline & \begin{matrix} 0\\0 \end{matrix} \\
  \hline 
     0 & \rvline & 0 \hspace{10mm}  0 & \rvline & 0 \\
\end{pmatrix}_{\alpha \beta}
\label{eq:adj4}
\end{align}
where ${\cal W}_{\alpha \beta}$ (${\cal W}_{\alpha \beta \gamma}$)  is the matrix formed by the $\alpha \beta $ ($\alpha \beta \gamma $)  rows and columns of $(2E)H$ with a=b=0, i.e. only the vacuum components.  Using $W_{\alpha \beta} \equiv \sum_i U_{\alpha i} U^*_{\beta i} m^2_i$,
\begin{align}
{\cal W}_{\mu,\tau} &=  \begin{pmatrix} 
W_{\mu \mu} & W_{\mu \tau} \\
W_{\tau \mu} & W_{\tau \tau} 
\end{pmatrix}  
\quad  \text{and} \quad {\cal W}_{\mu,\tau,s}   =  \begin{pmatrix} 
W_{\mu \mu} & W_{\mu \tau} & W_{\mu s} \\
W_{\tau \mu} & W_{\tau \tau} & W_{\tau s} \\
W_{s \mu} & W_{ s \tau} &W_{ss} 
\end{pmatrix}. 
\end{align}
Where $\text{\bf Adj}[{\cal W}_{\mu,\tau}]=\text{tr}({\cal W}_{\mu,\tau})I - {\cal W}_{\mu,\tau}$  and the elements of $\text{\bf Adj}[{\cal W}_{\mu,\tau,s}] $ are bilinear combinations of the W's and the identity 
\begin{align}
W_{\alpha \beta}W_{\gamma \delta} -W_{\alpha \delta }W_{\gamma \beta} &= \sum_{i>j} m^2_i m^2_j (U_{\alpha i} U_{\gamma j} -U_{\alpha j} U_{\gamma i} )(U_{\beta i} U_{\delta j}-U_{\beta j} U_{\delta i})^*
\end{align}
is useful. Note, if $\alpha = \gamma$ or  $\beta = \delta$ this bilinear sum is zero, so this combination never appears. Again see appendix \ref{sec:SUI}.

The eigenvectors are then calculated as follows:
\begin{align}
V_{\alpha i}V^*_{\beta i} & =\frac{\text{\bf Adj}[\lambda_i I_4 -(2E)H_{4x4}]_{\alpha \beta}}{\Pi_{j\neq i} (\lambda_i-\lambda_j)} \end{align}
with the numerator obtained from eq. \ref{eq:adj4} by making the $m^2_i \rightarrow (m^2_i-\lambda)$  replacement, which also implies a $W_{\alpha \beta} \rightarrow W_{\alpha \beta}-\lambda \delta_{\alpha \beta}$ replacement.
Defining  $$\ndelta_{\alpha \beta} \equiv 1- \delta_{\alpha \beta},$$ the opposite of the kronecker delta, we have
\begin{align}
\text{\bf Adj}[\lambda I_4-H_{4x4}]_{\alpha \beta} &= \lambda^3 \delta_{\alpha \beta}  \notag \\
& -\lambda^2 \biggr( \sum_i m^2_i( \delta_{\alpha \beta} -U_{\alpha i}U^*_{\beta i}) +a \, \ndelta_{e\alpha}
\, \delta_{\alpha \beta} +b \,\ndelta_{s\alpha}
 \,\delta_{\alpha \beta} \biggr)  \notag \\
&+\lambda \biggr( \sum_{i>j} m^2_i m^2_j( \delta_{\alpha \beta} -U_{\alpha i}U^*_{\beta i}-U_{\alpha j}U^*_{\beta j}) 
+a \, \ndelta_{e\alpha} \ndelta_{e\beta} \, ({\cal W}_{\mu,\tau,s}-\text{Tr}[{\cal W}_{\mu,\tau,s}] I_3)_{\alpha \beta}  \notag \\
& \hspace{2cm} + b \, \ndelta_{s\alpha}\ndelta_{s\beta} \, ({\cal W}_{e,\mu,\tau}-\text{Tr}[{\cal W}_{e,\mu,\tau}] I_3)_{\alpha \beta}  
+ ab  \, \ndelta_{e\alpha} \,\ndelta_{s\alpha}
 \, \delta_{\alpha \beta}
\biggr)  \notag \\
& - \biggr( \sum_i U_{\alpha i}U^*_{\beta i} ~ \Pi_{j \neq i}m^2_j  
  +a \, \ndelta_{e\alpha} \ndelta_{e\beta} \, \text{\bf Adj}[{\cal W}_{\mu,\tau,s}]_{\alpha \beta} 
   +b \,\ndelta_{s\alpha}\ndelta_{s\beta} \,\text{\bf Adj}[{\cal W}_{e,\mu,\tau}]_{\alpha \beta}  \notag \\
& \hspace{2cm}  
-ab\, \ndelta_{e\alpha} \ndelta_{e\beta} \, \ndelta_{s\alpha}\ndelta_{s\beta} \, ({\cal W}_{\mu,\tau}-\text{Tr}[{\cal W}_{\mu,\tau}] I_2)_{\alpha \beta} 
 \biggr) \, ,
\end{align}
where
\begin{align}
(\text{Tr}[{\cal W}_{\mu,\tau,s}] I_3-{\cal W}_{\mu,\tau,s})_{\alpha \beta}  &= \sum m^2_i (\delta_{\alpha \beta} (1-|U_{ei}|^2)-U_{\alpha i}U^*_{\beta i} ) \,,  \notag \\
(\text{Tr}[{\cal W}_{e,\mu,\tau}] I_3-{\cal W}_{e,\mu,\tau})_{\alpha \beta}  &= \sum m^2_i (\delta_{\alpha \beta} (1-|U_{si}|^2)-U_{\alpha i}U^*_{\beta i} ), \notag \\
(\text{Tr}[{\cal W}_{\mu,\tau}] I_2-{\cal W}_{\mu,\tau})_{\alpha \beta} &=\sum m^2_i (\delta_{\alpha \beta} (1-|U_{ei}|^2-|U_{si}|^2)-U_{\alpha i}U^*_{\beta i} ). \notag
\end{align}
Everything is now available for calculating the eigenvectors, $V_{\alpha i} V^*_{\beta i}$, as well as the oscillation probability using eq. \ref{eq:prob}.


\subsection{Matter potential invariant quantities for four flavors and beyond: }

As in the the 3x3 case, the CP conserving identities are given by
\begin{align}
\sum_{j>k} {\cal R}(V_{\alpha j}V^*_{\beta j} V^*_{\alpha k}V_{\beta k} )~ (\Delta \lambda_{jk})^2  & =  \sum_{j>k} {\cal R}(U_{\alpha j}U^*_{\beta j} U^*_{\alpha k}U_{\beta k} )~ (\Delta m^2_{jk})^2  \quad \alpha \neq \beta \,, \notag \\[2mm]
 \sum_{j>k} |V_{\alpha j}|^2 |V_{\alpha k}|^2 (\Delta \lambda_{jk})^2  & = \sum_{j>k} |U_{\alpha j}|^2 |U_{\alpha k}|^2 (\Delta m^2_{jk})^2
 \end{align}
 These identities are easily derived from the matter potential independence of $|H_{\alpha \beta}|^2$, $\alpha \neq \beta$ in the flavor basis, and are valid when there is more than one sterile neutrino.  In general these identities are valid whatever the matter potential is only along the diagonal in the flavor basis.  As we will see in the next section, non-diagonal entries in the flavor basis of matter potential to the Hamiltonian break these identities.
 \\

The CP violating identities can be derived using the matter potential independence of  
\begin{align}
\sum_{\gamma \neq \alpha, \beta} \Im(H_{\alpha \beta} H_{\beta \gamma} H_{\gamma \alpha} )
\end{align}
$\alpha \neq \beta$ and $\alpha, \beta, \gamma = e, \mu, \tau, s $. This is the generalization of what was done for 3 flavors to 4 (or more flavors).  This leads to a set of matter potential invariants that can be written as
\begin{align}
\sum_{i> j} \Im(V_{\alpha i} V^*_{\beta i} V^*_{\alpha j} V_{\beta j}) ~ \Delta \lambda_{ij}   \Delta \lambda_{ik}   \Delta \lambda_{jk} 
&=\sum_{i> j} \Im(U_{\alpha i} U^*_{\beta i} U^*_{\alpha j} U_{\beta j}) ~ \Delta m^2_{ij}   \Delta m^2_{ik}   \Delta m^2_{jk} 
\end{align}
for any fixed k. Note, when i=k or j=k, that term vanishes.\\

Using the notation\footnote{Some useful properties of the J's are  $J^{\alpha \beta}_{ij}= -J^{\alpha \beta}_{ji}=-J^{ \beta \alpha}_{ij}=J^{ \beta \alpha}_{ji}$ and $J^{\alpha \beta}_{ij}= -\sum_{k \neq j} J^{\alpha \beta}_{ik}=\sum_{k\neq i}J^{\alpha \beta}_{kj}$.
}
 $J^{\alpha \beta}_{ij}(U) \equiv \Im(U_{\alpha i} U^*_{\beta i} U^*_{\alpha j} U_{\beta j}) $, then for 4 flavors the CP violating piece of the appearance oscillation probability was given in Sec \ref{sec:oscprobs} as 
\begin{align}
 \sum_{i>j} J^{\alpha \beta}_{ij}(V) \sin 2 \Delta_{ij} 
 = & \quad 4  J^{\alpha \beta}_{32}(V) \sin\Delta_{32}  \sin\Delta_{21}  \sin\Delta_{13}   \notag  \\[-3mm]
&+ 4  J^{\alpha \beta}_{42}(V) \sin\Delta_{42}  \sin\Delta_{21}  \sin\Delta_{14}  \notag   \\[2mm]
&+ 4  J^{\alpha \beta}_{43}(V) \sin\Delta_{43}  \sin\Delta_{31}  \sin\Delta_{14}  
\end{align}
using k=1. 
In L/E $\rightarrow 0 $ limit, the coefficient of the $(L/E)^3$ terms is given by
\begin{align}
J^{\alpha \beta}_{32} (V) \, \Delta \lambda_{21}   \Delta \lambda_{31}   \Delta \lambda_{32} 
+ J^{\alpha \beta}_{42}(V) \,  \Delta \lambda_{21}   \Delta \lambda_{41}   \Delta \lambda_{42}  
+  J^{\alpha \beta}_{43} (V)  \,  \Delta \lambda_{41}   \Delta \lambda_{31}   \Delta \lambda_{43}
 \label{eq:4NHSinv}
\end{align}
and is therefore independent of matter potentials, i.e. is invariant.
 Such identities are the 4-flavor equivalents to the NHS ID.  There are 6 of them, but there only 3 independent ones: if one chooses the independent ones to be $(\alpha, \beta) = (e,\mu), (e,s) ~\&~ (\mu,s)$ then the others $(e, \tau), ( \mu, \tau) ~ \& ~ (\tau, s)$ are linear combinations as follows: since
\begin{align}
J^{e \tau}_{ij} =- J^{e \mu}_{ij}-J^{e s}_{ij}, \quad 
J^{\mu \tau}_{ij} =+J^{e \mu}_{ij}-J^{\mu s}_{ij}, \quad 
J^{\tau s}_{ij} =- J^{e s}_{ij}-J^{ \mu s}_{ij}   \, ,
\end{align}
 for $i\neq j$.
This is confirmed by the fact that there are 3 independent Dirac CP phases for 4 neutrino flavors.  All of these identities have been confirmed using the expressions from the previous sub-section.

For n-flavor of neutrinos, with n-3 sterile neutrinos, the equivalent NHS invariant  follows from the matter potential independence of 
$$\sum_{\gamma \neq \alpha, \beta} \Im(H_{\alpha \beta} H_{\beta \gamma} H_{\gamma \alpha} ) .$$
The identities are 
\begin{align}
 \sum_{i>j} \Delta m^2_{ij}  \Delta m^2_{jk} \Delta m^2_{ki} ~ J^{\alpha \beta}_{ij}(U) &= \sum_{i>j}    \Delta \lambda_{ij}  \Delta \lambda_{jk}   \Delta \lambda_{ki}  \, J^{\alpha \beta}_{ij} (V)
 \label{eq:nNHSinv}
\end{align}
for k=(1, 2, $\cdots$, n) and $\alpha, \beta = (e, \mu, \tau, s_1, \cdots) $.  Although there are a large number of identities here only  (n-1)(n-2)/2  are independent.  A proof of this identity is given in appendix \ref{sec:nNHS}. Even for one sterile neutrino eq. \ref{eq:nNHSinv} is more general than eq. \ref{eq:4NHSinv}, since k can take any value from 1 to 4.


\section{3 flavors with off-diagonal non-standard interactions (NSI)}
\label{sec:3f-nsi}

Here we sketch how the systematic methods presented earlier can be used for the case when there are non-standard interactions in the off-diagonal components of the Hamiltonian in the flavor basis\footnote{For non-standard interactions in the diagonal components the method advocated here is also simple to apply.}.

\subsection{NSI e$\mu$}
The Hamiltonian is 
\begin{align}
(2E) H &=U M^2 U^\dagger 
+
\left(\begin{array}{ccc} 
a& b& 0\\
b^*&0&0\\
0 & 0 &0
\end{array}
\right)\, ,
\end{align}
where in the usual notation $b \equiv a \, \epsilon_{e \mu}$ in this subsection.
The determinant is given by
\begin{align}
&\text{\bf Det}[(2E)H] -\text{\bf Det}[(2E)H_{3x3}] \notag \\
&=  -b(W_{\mu e}W_{\tau \tau}-W_{\mu \tau}W_{\tau e})-b^*(W_{e \mu}W_{\tau \tau}-W_{\tau \mu}W_{e \tau })-bb^*W_{\tau \tau} \,, \notag \\
&=
 \sum_{P} m^2_i m^2_j (b \,U^*_{ek}U_{\mu k} +b^*\, U_{ek}U^*_{\mu k}) -bb^* \sum_i m^2_i |U_{\tau i}|^2
\end{align}
where $\text{\bf Det}[(2E)H_{3x3}]$ is given by eq. \ref{eq:detH3x3}. Then the characteristic equation can be calculated using the replacement $m^2_i \rightarrow (m^2_i -\lambda)$, giving
\begin{align}
\text{\bf Det}[\lambda I-(2E)H]&=  \text{\bf Det}[\lambda I-(2E)H_{3x3}] -  \lambda (bW_{\mu e} +b^* W_{e \mu})  +bb^* W_{\tau \tau} =0 \,,
\end{align}
where $\text{\bf Det}[\lambda I-(2E)H_{3x3}]$ is given by eq. \ref{eq-det3}.

The eigenvectors can be determined from $\text{\bf Adj}[(2E)H] $ in a similar fashion.  First,
\begin{align}
& \text{\bf Adj}[(2E)H] - \text{\bf Adj}[(2E)H_{3x3}] \notag \\
& \quad = - b \begin{pmatrix}
0 & W_{\tau \tau} & -W_{\mu \tau} \\
0 & 0 & 0 \\
0& -W_{\tau e } & W_{\mu e } 
\end{pmatrix}
 \notag 
 -b^* \begin{pmatrix}
0 & 0 & 0\\
W_{\tau \tau} &  0&  -W_{e \tau} \\
-W_{\tau \mu} & 0 & W_{e \mu } 
\end{pmatrix}
 - b b^* \begin{pmatrix}
0 & 0 & 0\\
0 & 0 & 0 \\
0 & 0 & 1
\end{pmatrix} 
\end{align}
where $\text{\bf Adj}[(2E)H_{3x3}]$ is given by eq. \ref{eq:adjH3x3}. Then by using the  $m^2_i \rightarrow (m^2_i -\lambda)$ replacement, we have
\begin{align}
& \text{ \bf Adj}[\lambda I-(2E)H]  \notag \\
&. \quad  =\text{\bf Adj}[\lambda I- (2E)H_{3x3}]+ \lambda  \begin{pmatrix}
0 & b & 0 \\
b^* & 0 & 0 \\
0& 0 & 0
\end{pmatrix} +( \text{\bf Adj}[(2E)H] - \text{\bf Adj}[(2E)H_{3x3}] )
\end{align}
where $\text{\bf Adj}[(2E)H_{3x3}]$ and  $\text{\bf Adj}[\lambda I-(2E)H_{3x3}]$ are given by eq. \ref{eq:adjH3x3} and \ref{eq-Adj3}, respectively.

Since, in the flavor basis, $H_{e \tau}$ and $H_{\mu \tau}$ are independent of the matter potential, therefore we have
\begin{align}
\sum_{j>k} {\cal R}(V_{\alpha j}V^*_{\beta j} V^*_{\alpha k}V_{\beta k} )~ (\Delta \lambda_{jk})^2  & =  \sum_{j>k} {\cal R}(U_{\alpha j}U^*_{\beta j} U^*_{\alpha k}U_{\beta k} )~ (\Delta m^2_{jk})^2
\end{align}
is invariant for $(\alpha, \beta) = (e, \tau), (\tau, e), (\mu,  \tau) ~ \text{and}  (\mu,  \tau)$ but {\it not} $(e, \mu)$ or $(\mu, e)$.  Adding  $(\tau, e)$ and $(\tau, \mu)$ one also obtains 
\begin{align}
 \sum_{j>k} |V_{\tau j}|^2 |V_{\tau k}|^2 (\Delta \lambda_{jk})^2  & = \sum_{j>k} |U_{\tau j}|^2 |U_{\tau k}|^2 (\Delta m^2_{jk})^2
 \end{align}
as an invariant. So in the small L/E, the $(L/E)^2$ coefficients of the CP conserving part of the oscillation probabilities are independent of the matter potential for $\nu_e \leftrightarrow \nu_\tau$,  $\nu_\mu \leftrightarrow \nu_\tau$ and $\nu_\tau$ disappearance but not for $\nu_e \leftrightarrow \nu_\mu$ or $\nu_e$ and $\nu_\mu$ disappearance.\\

For the equivalent of the  NHS identity, or CPV invariant,  the dependence of $H_{e \mu}$ on the matter potential means that
$\Im(H_{e \mu} H_{\mu \tau} H_{\tau e})$ also depends on the matter potential and therefore in the small L/E limit, the coefficient $(L/E)^3$ term of the CP violating part of the appearance oscillation probabilities depends on the matter potential for all channels, see also \cite{Fong:2022oim}. This statement is true if the off-diagonal matter effect appears in $H_{e \tau}$ or $H_{\mu \tau}$. So any off-diagonal contribution to the matter potential in the flavor basis renders the NHS expression matter potential dependent.\\

Similar results can be obtained for an NSI in the e$\tau$ sector.

\subsection{NSI $\mu \tau$}
For an NSI component in the  $\mu \tau$ sector, the Hamiltonian is given by
\begin{align}
(2E) H &=U M^2 U^\dagger +
\left(\begin{array}{ccc} 
a& 0&0\\
0&0&b\\
0 & b^* &0
\end{array}
\right) \,,
\end{align}
where in the usual notation $b \equiv a\epsilon_{\mu \tau}$ in this subsection.
 The determinant and adjugate of the Hamiltonian are slightly different as product terms between ``a'' and ``b'' appear:
\begin{align}
\text{\bf Det}[(2E)H] 
&=  \text{\bf Det}[(2E)H_{3x3}] \notag \\ & -b(W_{ee} W_{\tau \mu}-W_{e \mu}W_{\tau e})-b^*(W_{ee}W_{\mu \tau}-W_{\mu e}W_{e \tau}) \notag \\
& - a(bW_{\tau \mu}+b^*W_{\mu \tau}) -bb^*(a+W_{ee})
\end{align}
where $\text{\bf Det}[(2E)H_{3x3}]$ is given by eq. \ref{eq:detH3x3}. Similarly
\begin{align}
\text{\bf Adj}[(2E)H] &= \text{\bf Adj}[(2E)H_{3x3}] \notag \\[3mm]
& \hspace*{-2cm}
 -b \begin{pmatrix}
 W_{\tau \mu} & 0 & -W_{e \mu} \\
 -W_{\tau  e } & 0 & W_{e e} \\
0 & 0 & 0
\end{pmatrix}
-b^* \begin{pmatrix}
 W_{\mu \tau} & -W_{e \tau}&0 \\
 0 & 0 & 0\\
 -W_{\mu e} & W_{e e} &0 
\end{pmatrix}
-a \begin{pmatrix}
0 & 0 & 0\\
0 & 0 & b \\
0 & b^* & 0
\end{pmatrix}
 -b b^* \begin{pmatrix}
1 & 0 & 0\\
0 & 0 & 0 \\
0 & 0 & 0
\end{pmatrix}  \,.
\end{align}
Both $\text{\bf Det}[\lambda I-(2E)H]$ and $\text{\bf Adj}[\lambda I-(2E)H]$ can be calculated using the
$m^2_i \rightarrow (m^2_i -\lambda)$ procedure as before.\\

Using similar arguments as in $(e, \mu)$ case, in the small L/E, the $(L/E)^2$ coefficients of the CP conserving part of the oscillation probabilities are independent of the matter potential for $\nu_e \leftrightarrow \nu_\mu$,  $\nu_e \leftrightarrow \nu_\tau$ and $\nu_e$ disappearance but not for $\nu_\mu \leftrightarrow \nu_\tau$ or $\nu_\mu$ and $\nu_\tau$ disappearance.\\

If the NSI components appear in more than one off-diagonal component of the Hamiltonian the calculation of $\text{\bf Det}[\lambda I-(2E)H]$ and $\text{\bf Adj}[\lambda I-(2E)H]$  follows in a similar fashion but of course the resulting equations are more complicated.  For this case it maybe simpler instead of using $\text{\bf Adj}[\lambda_i I-(2E)H]$ in the numerator of eq. \ref{eq-Adj} but to use ${\bf \Pi}_{j\neq i} (H-\lambda_j I)$ as for three neutrinos only $H$ and $H^2$ appear.



\section{Conclusion}
\label{sec:concl}

Neutrino oscillations in matter is important for measuring the remaining unknown parameters in the neutrino sector as well as for the discovery of beyond the Standard Model physics in this sector.
The LBL experiments in particular will discover the final piece of the neutrino mass ordering, $\Delta m^2_{32} >\text{or}< 0$, determine whether the neutrino mass state with least $\nu_e$ is dominated by $\nu_\mu$ or $\nu_\tau$, octant of $\theta_{23}$, and determine the size and sign of CP violation in the neutrino sector, i.e. determine the Jarlskog invariant. All of the current and future LBL experiments, T2K, NOvA, HK, DUNE and KNO, have significant matter effects which must be included in their analysis.  Understanding these matter effects for the Standard Model three flavor paradigm as well as for 
beyond the Standard Model scenarios is important for extracting the most physics out of these expensive experiments.\\

In this paper we have provided a new simple way to determine the mixing matrix in constant density matter once the mass eigenvalues have been determined. This method uses a known result in linear algebra that has not been used in the neutrino literature before and has general applicability in particle physics. We call this method the adjugate method as it uses the adjugate of the Hamiltonian to determine the mixing matrix in matter. This method is particularly well suited to neutrino oscillations due to the form of the Hamiltonian but is also very general and can used for any Hamiltonian.  Once the mixing matrix in matter has been determined it is then very simple to determine the oscillation probabilities for these LBL experiments. We have also rewritten the  CP violating part of the neutrino oscillation probability in a way that better reflects the physics of this important part of the oscillation probability, see Sec. \ref{sec:oscprobs}, and is especially useful in our discussion of four or more flavor neutrino oscillations.\\

In Sec. \ref{sec:3f}, we have revisited matter effects for three flavor oscillations that has previously been addressed by many authors.  The derivation here is particularly simple and straight forward such that we do not need the form of the PMNS matrix to obtain results.  We have also derived all of the useful matter invariant quantities including the NHS identity eq. \ref{eq:nhs} and the KTY identity eq.~\ref{eq:kty} as well as identities related to the disappearance probabilities eq.~\ref{eq:3disapp} and the CPC part of the appearance probabilities, eq. \ref{eq:3CPC}. The scenario with three active and one sterile is addressed in Sec. \ref{sec:4f}. Again we derive simple expressions for the mixing matrix in matter and give a number of invariant quantities especially the generalization of the NHS identity to four or more neutrinos. This is a sum of three neutrino factors which reduces to three flavor result as the fourth neutrino is decoupled.
In Sec. \ref{sec:3f-nsi}, we have given a sketch of how to apply this method to Hamiltonians with NSI and discuss what happens to the invariant quantities with different non-zero off-diagonal elements in the Hamiltonian in the flavor basis.\\

The adjugate method for determining the mixing matrix, i.e. the eigenvectors of the Hamiltonian,  is of general applicability but is especially well suited to determining neutrino oscillation probabilities in constant density matter due to the form of the Hamiltonian. We have illuminated its use in a variety of scenarios from the three flavor Standard Model as well as additional sterile neutrinos and also with non-standard interactions. We have used this method to illuminate a variety of quantities that are independent of the matter potential in the various scenarios and discussed the physics of these invariants.
Extension of the adjugate method to non-normal Hamiltonians will be the subject of a follow up paper.

\acknowledgments
We are very grateful to Peter Denton for discussions, encouragement and comments on this manuscript.
Fermilab is operated by the Fermi Research Alliance under contract no.~DE-AC02-07CH11359 with the U.S.~Department of Energy.  This project was initiated at the KITP's program,  ``Neutrinos as a Portal to New Physics and Astrophysics'', and was supported in part by the National Science Foundation under Grant No. PHY-1748958.
This project has received funding/support from the European Union's Horizon 2020 research and innovation programme under the Marie Sklodowska-Curie grant agreement No 860881-HIDDeN.

\newpage

\appendix


\section{Rewriting the CP violating term in Oscillation Probabilities}
\label{sec:CPVterm}.

First we defining $J^{\alpha \beta}_{ij}$ and note some of J's properties:
\begin{align}
 J^{\alpha \beta}_{ij} &\equiv {\Im}( V_{\alpha i} V^*_{\beta i}   V^*_{\alpha j}  V_{\beta j} ) \notag \\
 J^{\alpha \beta}_{ij} &= -J^{\alpha \beta}_{ji}=-J^{ \beta \alpha}_{ij}=J^{ \beta \alpha}_{ji} \notag \\
 J^{\alpha \beta}_{kj}  &= -\sum_{i\neq k} J^{\alpha \beta}_{ij}, \quad 
 J^{\alpha \beta}_{ik}  = -\sum_{j\neq k} J^{\alpha \beta}_{ij}, \quad  J^{\alpha \beta}_{kk}  =0 \, .
 \end{align}
 Using the above identities one can show that
 \begin{align}
\sum_{i,j} J^{\alpha \beta}_{ij} \sin 2 \Delta_{ij} & = \sum_{i \neq k, j \neq k } J^{\alpha \beta}_{ij}  \sin 2 \Delta_{ij}+\sum_{ j \neq k } J^{\alpha \beta}_{kj}  \sin 2 \Delta_{kj} + \sum_{i \neq k } J^{\alpha \beta}_{ik}  \sin 2 \Delta_{ik} + J^{\alpha \beta}_{kk}  \sin 2 \Delta_{kk} \notag  \\ 
  \sum_{i,j} J^{\alpha \beta}_{ij} \sin 2 \Delta_{ij} & = \sum_{i \neq k, j \neq k } J^{\alpha \beta}_{ij} \biggr( \sin 2 \Delta_{ij}+ \sin 2 \Delta_{jk}+ \sin 2 \Delta_{ki}\biggr) \notag \\
  &=4\sum_{i,j } J^{\alpha \beta}_{ij} \biggr( \sin\Delta_{ij}\sin\Delta_{jk}\sin\Delta_{ik} \biggr) \, .
\end{align}
For the last line the trig identity 
\begin{align}
\sin 2 \Delta_{ij} +\sin 2 \Delta_{jk}+ \sin 2 \Delta_{ki} = - 4 \sin \Delta_{ij} \sin \Delta_{jk} \sin \Delta_{ki}
\end{align}
was used. Therefore, for any fixed k
 \begin{align}
\sum_{i>j} J^{\alpha \beta}_{ij} \sin 2 \Delta_{ij} & =
 4\sum_{i>j } J^{\alpha \beta}_{ij} \biggr( \sin\Delta_{ij}\sin\Delta_{jk}\sin\Delta_{ik} \biggr) \, .
\end{align}
This is eq. \ref{eq:CPV}.


\section{Adjugate ``Proof''}
\label{sec:adjproof}

To demonstrate
\begin{align}
V_{\alpha i} V^*_{\beta i} &= \frac{\text{\bf Adj}[\lambda_i I -H]_{\alpha \beta}}{\Pi_{j\neq i} (\lambda_i -\lambda_j)} \, .
\label{eq:AdjB}
\end{align}
Consider the case when $H=V\Lambda V^\dagger$ with $\Lambda=\text{\bf Diag}(\lambda_1, \lambda_2, \cdots, \lambda_n)$,  then 
\begin{align}
\text{\bf Adj}[\lambda_i I -H]_{\alpha \beta}  &= V_{\alpha j} \, \text{\bf Adj}[\lambda_i I -\Lambda]_{jk} \,V_{\beta k}^*  \notag \\
&=\Pi_{j\neq i} (\lambda_i -\lambda_j)   V_{\alpha i}  \,V_{\beta i}^* \,  
\end{align}
where we have used $\text{\bf Adj}[UAU^\dagger]= \text{\bf Adj}[U^\dagger]  \,  \text{\bf Adj}[A]  \, \text{\bf Adj}[U]=U\, \text{\bf Adj}[A]\, U^\dagger$, when $U$ is unitary matrix. qed.

Alternatively, eq. \ref{eq:AdjB} can be rewritten as follows:
\begin{align}
V_{\alpha i} V^*_{\beta i}  &= \lim_{\lambda \rightarrow \lambda_i} \, \left( \frac{\text{\bf Adj}[\lambda I -H]_{\alpha \beta}}{\Pi_{j\neq i} (\lambda -\lambda_j)}  \right) 
=  \lim_{\lambda \rightarrow \lambda_i} \,  \left( \frac{\text{\bf Det}[\lambda I-H] \, (\lambda I -H)^{-1}_{\alpha \beta}}{\Pi_{j\neq i} (\lambda -\lambda_j)} \right)  \, .
\end{align}
Here, one needs to consider the limit $ {\lambda \rightarrow \lambda_i}$ as the matrix $(\lambda_i I -H)$ is not invertible as \\ $\text{\bf Det}[\lambda_i I-H] =0$, although the $ \text{\bf Adj}[\lambda_i I -H]$ is well defined.
Now consider the case when $H=V\Lambda V^\dagger$  $(H^{-1} = V \Lambda^{-1} V^\dagger)$ with $\Lambda=\text{\bf Diag}(\lambda_1, \lambda_2, \cdots, \lambda_n)$, so that
$$ \text{\bf Det}[\lambda I-H] = \Pi_j (\lambda-\lambda_j) \quad  \text{and} \quad   (\lambda I -H)^{-1}_{\alpha \beta} =\sum_k V_{\alpha k} \frac{1}{(\lambda-\lambda_k)} V^*_{\beta k} ~.$$
Then
\begin{align}
 \frac{\text{\bf Adj}[\lambda_i I -H]_{\alpha \beta}}{\Pi_{j\neq i} (\lambda_i -\lambda_j)}  &=  \lim_{\lambda \rightarrow \lambda_i} \, \left(  \frac{\Pi_j (\lambda-\lambda_j)  \,\sum_k V_{\alpha k} \frac{1}{(\lambda-\lambda_k)} V^*_{\beta k}}{\Pi_{j\neq i} (\lambda -\lambda_j)}  \right) \notag \\
 &= \lim_{\lambda \rightarrow \lambda_i} \, \left( \sum_k V_{\alpha k} \frac{(\lambda-\lambda_i)}{(\lambda-\lambda_k)} V^*_{\beta k} \right)  \notag \\
 &= V_{\alpha i} V^*_{\beta i}
 \end{align}
 qed
 
 For any matrix, X, the following identity holds:
 \begin{align}
 X.\text{\bf Adj}[X]=\text{\bf Adj}[X].X=\text{\bf Det}[X].I \, .
 \end{align}
 Thus by setting $X=\lambda_i I-H$, so that RHS is zero, demonstrates that $$ \text{\bf Adj}[\lambda_i I-H]_{\alpha \beta} \propto
 V_{\alpha i} V^*_{\beta i},$$ since $HV=V\Lambda$ and $V^\dagger H = \Lambda V^\dagger$. 
 Since we require $\sum_{\alpha} |V_{\alpha i}|^2 =1$, the normalization is obtained using 
 $$\text{\bf Tr}[ \text{\bf Adj}[\lambda_i I-H]]= {\bf \Pi}_{j\neq i} (\lambda_i -\lambda_j)$$ giving eq. \ref{eq:AdjB}.


\section{Le\,Verrier-Faddeev algorithm.}
\label{app:LeVF}
Following ref. \cite{Hou:1998jdv}, we can write for any $\lambda$ the following:
\begin{align}
\text{Det}(\lambda I-H) &= \lambda^{n} + \lambda^{n-1 }d_1+\lambda^{n-2} d_2+\cdots+d_n \, , \\
\text{Adj}(\lambda I-H) &\equiv \lambda^{n-1} A_1+\lambda^{n-2} A_2+\cdots+A_n \,,
\end{align}
where the A's and d's are calculated recursively as follows
\begin{align}
A_1&=I \,, \quad d_1= -\frac1{1}\text{Tr}(HA_1)=-\text{Tr}(H) \\
A_2&=HA_1+d_1I = H-\text{Tr}(H)I \notag  \\
d_2 &= - \frac1{2} \text{Tr}(HA_2)=\frac1{2}(\text{Tr}^2(H) -\text{Tr}(H^2))  \,.
\end{align}
If H is 2x2 matrix then $A_2 = -\text{Adj}(H)$ and $d_2=\text{Det}(H)$.
Continuing
\begin{align}
A_3&=HA_2+d_2I =H^2+d_1 H + d_2 I 
 \,, \notag \\
d_3 & = - \frac1{3} \text{Tr}(HA_3) = - \frac1{6} ( \text{Tr}^3(H)- 3 \text{Tr}(H)\text{Tr}(H^2)+2 \text{Tr}(H^3)).
\end{align}
If H is a 2x2 matrix, $A_3=0$ and $d_3=0$. For 3x3 matrix then $A_3 = \text{Adj}(H)$ and $d_3=-\text{Det}(H)$. Next iteration gives
\begin{align}
A_4 &=HA_3+d_3I  = H^3 +d_1 H^2 + d_2 H +d_3 I \notag
\, , \\
d_4 
&
= \frac1{24} ( \text{Tr}^4(H)-
 6 \text{Tr}^2(H)\text{Tr}(H^2)+3 \text{Tr}^2(H^2) +8 \text{Tr}(H)\text{Tr}(H^3) -6 \text{Tr}(H^4))
 \, .
 \end{align}
If H is a 3x3 matrix or smaller, $A_4=0$ and $d_4=0$. For  4x4 matrix then $A_4 =- \text{Adj}(H)$ and $d_4=\text{Det}(H)$. Then,
\begin{align}
A_5&=HA_4+d_4I =H^4+d_1 H^3 + d_2 H^2 + d_3 H +d_4 I \notag
  \, ,  \\
  d_5 
&= -\frac1{120} [ ~\text{Tr}^5(H)-
 10 \text{Tr}^3(H)\text{Tr}(H^2)+15 \text{Tr}(H) \text{Tr}^2(H^2) +20 \text{Tr}^2(H)\text{Tr}(H^3) \notag  \\
 & \hspace{+3cm} -20 \text{Tr}(H^2) \text{Tr}(H^3)-30 \text{Tr}(H)\text{Tr}(H^4)+24 \text{Tr}(H^5) ~]. 
  \end{align}
If H is a 4x4 matrix or smaller, $A_5=0$ and $d_5=0$. For 5x5 matrix then $A_5 =\text{Adj}(H)$ and $d_5=-\text{Det}(H)$. Next iteration gives
\begin{align}
 A_6&=HA_5+d_5I = =H^5+d_1 H^4 + d_2 H^3 + d_3 H^2 +d_4 H +d_5 I  \notag \\
d_6 
&=
 \frac1{720} [ ~\text{Tr}^6(H)-
 15~ \text{Tr}^4(H)\text{Tr}(H^2)+45 ~\text{Tr}^2(H) \text{Tr}^2(H^2) -15~ \text{Tr}^3(H^2)  \notag  \\
 &  \quad \quad  +40~ \text{Tr}^3(H) \text{Tr}(H^3) -120~ \text{Tr}(H)\text{Tr}(H^2)\text{Tr}(H^3)+40~ \text{Tr}^2(H^3) 
   \\
 & \quad \quad  -90~ \text{Tr}^2(H)\text{Tr}(H^4) +90~ \text{Tr}(H^2)\text{Tr}(H^4)  +144~ \text{Tr}(H)\text{Tr}(H^5)-120 ~\text{Tr}(H^6)~]  \notag
 \end{align}
If H is a 5x5 matrix or smaller, $A_6=0$ and $d_6=0$. For  6x6 matrix then $A_6 =- \text{Adj}(H)$ and $d_6=\text{Det}(H)$.  All of the above satisfy $\text{\bf Tr}[A_{m+1}] =(n-m)d_m$ as required.

For an nxn matrix this continues until 
\begin{align}
A_n&=HA_{(n-1)}+d_{(n-1)}I \notag \\
& =\text{Adj}(-H)=(-1)^{n-1} \text{Adj}(H) \label{eq:adjN}  \\
d_n &= - \frac1{n} \text{Tr}(HA_{n}) = (-1)^n \text{Det}(H) \, .
\end{align}
The next iteration returns zero for both $A_{n+1}$ and $d_{n+1}$, as follows
\begin{align}
A_{n+1}&= HA_n+d_n I =(-1)^{n-1}\biggr( H\text{Adj}(H)-\text{Det}(H)I \biggr)=0 \\
d_{n+1}&= - \frac1{n+1} \text{Tr}(HA_{n+1}) = 0 \,  \notag
\end{align}
thus terminating the series.\footnote{Note for any $d_k$ if you put all traces equal to ``k'', then $d_k=(-1)^k$.
This is equivalent to replacing H with $I_k$, and is a useful but simple cross check.}

This recursion method is simple to program for numerical studies.  Also eq. \ref{eq:adjN} gives an alternative way to calculate the $ \text{Adj}(H) $ for an nxn matrix in terms of powers of H up to (n-1) plus their traces, this is equivalent to the Caley-Hamilton identity.

It is worth noting that eq.   can be written as follows:
\begin{align}
\text{\bf Det}(\lambda I-H) &= \lambda^{n} - \lambda^{n-1 }\text{\bf Det}_1(H) +\lambda^{n-2} \text{\bf Det}_2(H) +\cdots+ (-1)^n \text{\bf Det}_n(H)  \, , \\
\text{\bf Adj}(\lambda I-H) &= \lambda^{n-1} I-\lambda^{n-2} \text{\bf Adj}_2(H)+\lambda^{n-3} \text{\bf Adj}_3(H) +\cdots+(-1)^{n-1} \text{\bf Adj}_n(H) \,,
\end{align}
where $\text{\bf Det}_m(H) $ and $\text{\bf Adj}_m(H)$ are defined as the polynomials of $H$, that gives the determinant of H and the adjugate of $H$ is if $H$ where an m\,x\,m matrix, i.e.
\begin{align}
\text{\bf Det}_1(H) &\equiv  \text{\bf Tr}(H) =\sum_j \lambda_j \ \notag   \\
\text{\bf Det}_2(H) &\equiv  \frac1{2}(\text{\bf Tr}^2(H)-\text{\bf Tr}(H^2))=\sum_{j1>j2} \lambda_{j1} \lambda_{j2} \notag \\
\text{\bf Det}_3(H) &\equiv  \frac1{24}(\text{\bf Tr}^3(H)-3\text{\bf Tr}(H)\text{\bf Tr}(H^2)+2 \text{\bf Tr}(H^3))=\sum_{j1>j2>j3} \lambda_{j1} \lambda_{j2} \lambda_{j3} \notag \\[-3mm]
\vdots \quad  & \quad  \quad  \vdots \notag \\
\text{\bf Det}_n(H) &\equiv \text{\bf Det}(H) = \Pi_j \lambda_j
\end{align}
and
\begin{align}
\text{\bf Adj}_2(H) &\equiv  \text{\bf Tr}(H)I-H \notag   \\
\text{\bf Adj}_3(H) &\equiv H^2 -\text{\bf Tr}(H)H + \frac1{2}(\text{\bf Tr}^2(H)-\text{\bf Tr}(H^2))I \notag \\
\vdots \quad  & \quad  \quad  \vdots \notag \\
\text{\bf Adj}_n(H) &\equiv \text{\bf Adj}(H) 
\end{align}


\section{Proof of $\text{Adj}(\lambda_i I-H) = {\bf \Pi}_{j\neq i} (H-\lambda_j I)$}
 \label{app:LeVF2} 

Rearrange the LeVerrier-Faddeev expression for $\text{\bf Adj}(\lambda I-H)$ as a power series in the matrices $I,~H, ~H^2, \cdots, ~H^{n-1}$:
\begin{align}
\text{\bf Adj}(\lambda I-H) &\equiv \lambda^{n-1} A_1+\lambda^{n-2} A_2+\cdots+A_n \,,  \notag \\
&=  \lambda^{n-1} I + \lambda^{n-2} (H+d_1 I) + \lambda^{n-3} (H^2+d_1 H +d_2 I)  \notag \\
& \quad + \cdots + (H^{n-1}+d_1H^{n-2} + \cdots +d_{n-1} I) \, .  \notag
\end{align}
Grouping powers of $H$, we have
\begin{align}
\text{\bf Adj}(\lambda I-H)  &= ( \lambda^{n-1} +d_1  \lambda^{n-2} + d_2  \lambda^{n-3}+ \cdots +d_{n-1}) I \notag \\
&\quad  +( \lambda^{n-2} + d_1  \lambda^{n-3} + d_2  \lambda^{n-4}+ \cdots + d_{n-2}) 
H \notag\\
& \quad \quad \quad \vdots \notag \\
&\quad  +( \lambda^2 + d_1\lambda +d_2)  H^{n-3} \notag\\
&\quad  +( \lambda + d_1)  H^{n-2} \notag\\
&\quad  +H^{n-1} 
\end{align}
Now for nxn Hamiltonian with eigenvalues $\lambda_j$, its easy to show that $$d_m =  (-1)^m \sum_{ j1>...>jm } \lambda_{j1} \cdots \lambda_{jm}$$ as these are just coefficients of the characteristic equation $\text{\bf Det}(\lambda I-H)= \Pi_i(\lambda-\lambda_i)=0$.

If $\lambda_i$ is one of the eigenvalues of $H$, $(\lambda_1, \cdots, \lambda_n)$, then one can show by induction that
\begin{align}
( \lambda_i +d_1)  &=- \sum_{ j1 \neq i }\lambda_{j1} \\
( \lambda^2_i +d_1\lambda_i +d_2)  &= \sum_{\begin{array}{c} j1>j2 \\[-1mm] \neq i \end{array} }\lambda_{j1} \lambda_{j2} \\
( \lambda^3_i + d_1\lambda^2_i +d_2 \lambda_i +d_3) & = - \hspace{-1cm}. \sum_{\begin{array}{c} j1>j2>j3 \\[-1mm] \neq i \end{array} }\lambda_{j1} \lambda_{j2} \lambda_{j3} \\
& \quad  \vdots \notag \\
( \lambda^{n-1}_i + d_1  \lambda^{n-2}_i + \cdots +d_{n-1}) &= (-1)^{n-1} \hspace{-1cm}
\sum_{\begin{array}{c} j1>...>j(n-1) \\[-1mm] \neq i \end{array} }\lambda_{j1} \cdots  \lambda_{j(n-1)}   \notag \\
&=   (-1)^{n-1}  \,  \Pi_{j\neq i}\lambda_j   \,, 
\end{align}
where $\lambda_i$ does not appear on the RHS of any of these identities. The following identity is use repetitively:
\begin{align}
 \sum_{\begin{array}{c} j1>...>jm \\[-1mm] \neq i \end{array} } \hspace{-0.5cm}  \lambda_{j1} \cdots  \lambda_{jm} 
& =  \sum_{\begin{array}{c} j1>...>jm \\[-1mm] \end{array} } \hspace{-0.5cm}  \lambda_{j1} \cdots  \lambda_{jm}  
\hspace{5mm}   - ~~ \lambda_i \hspace{-1cm}  \sum_{\begin{array}{c} j1>...>j(m-1) \\[-1mm] \neq i \end{array} } \hspace{-1cm}  \lambda_{j1} \cdots  \lambda_{j(m-1)} \,.
\end{align}

Therefore, using the fact that all the $\lambda$'s are eigenvalues of H, we have that
\begin{align}
\text{\bf Adj}(\lambda_i I-H) 
&= H^{n-1} -  H^{n-2}  \sum_{ j \neq i }\lambda_{j} 
+ H^{n-3} \hspace{-3mm} \sum_{\begin{array}{c} j1>j2 \\[-1mm] \neq i \end{array} }\lambda_{j1} \lambda_{j2} \notag \\ 
&  +\cdots+  (-1)^{n-1} \, I_n \,  \Pi_{j\neq i}\lambda_j  \,,  \notag \\
&={\bf  \Pi}_{j\neq i} (H-\lambda_j I) \,.
\end{align} 
 This expression can also be derived from the \href{https://proofwiki.org/wiki/Inverse_of_Vandermonde_Matrix}{inverse of the Vandermonde matrix}\,\footnote{https://proofwiki.org/wiki/Inverse\_of\_Vandermonde\_Matrix}. 
 
 Note, the  LeVerrier-Faddeev recursion method gives both $\text{\bf Adj}(\lambda I-H) $ and $\text{\bf Det}(\lambda I-H) $ for arbitrary $\lambda$, where as $\text{\bf Adj}(\lambda_i I-H) ={\bf  \Pi}_{j\neq i} (H-\lambda_j I) $ requires all of $\{\lambda_1, \cdots, \lambda_n\}$ to be eigenvalues of H. Both calculations require (n-1) matrix multiplications, so are of similar computational cost.  However, the LeVerrier-Faddeev recursion method automates the calculation of $\text{\bf Adj}(\lambda I-H) $ and $\text{\bf Det}(\lambda I-H) $ and thus require very few lines of code to program.

As a cross check consider the case that $H=V\Lambda V^\dagger$, then
\begin{align}
\Pi_{j \neq i} (H-\lambda_j I)|_{\alpha \beta} &= V~\Pi_{j \neq i} (\Lambda-\lambda_j I )~ V^\dagger |_{\alpha \beta} 
 =\sum_k V_{\alpha k} ~ \Pi_{j \neq i} (\lambda_k  -\lambda_j)~ V^*_{\beta k} \notag \\
&=V_{\alpha i} V^*_{\beta i} ~ \Pi_{j \neq i} (\lambda_i  -\lambda_j) \, ,
\end{align}
as required.\\

As an observation note that
\begin{align}
\Pi_{j \neq i} (H-\lambda_j I) &=H^{n-1} - \biggr( \sum_{j \neq i} \lambda_j \biggr) H^{n-2} + \cdots+(-1)^{n-2} \biggr(\lambda_{j1} \lambda_{j2} \cdots \lambda_{j(n-1)}\biggr)I. \\
\Pi_{j \neq i} (\lambda_i- \lambda_j) &=\lambda_i^{n-1} - \biggr( \sum_{j \neq i} \lambda_j \biggr) \lambda^{n-2}_i + \cdots+(-1)^{n-2} \biggr(\lambda_{j1} \lambda_{j2} \cdots \lambda_{j(n-1)}\biggr)\, ,\end{align}
where none of the $j$'s are equal i and the quantites in $(\cdots)$ are the \href{https://en.wikipedia.org/wiki/Elementary_symmetric_polynomial}{elementary symmetric polynomials} of the (n-1) variables $(\lambda_1, \cdots \lambda_n)$ that do not including $\lambda_i$,  which are sometimes written as $e_k(\{\lambda_1, \cdots,\lambda_n\}\backslash  \lambda_i)$ for $k=1, \cdots, n-1$.  Both the numerator and denominator 
of eq. \ref{eq:Adj.eq.Pi} are identical polynomials of $H$ and $\lambda_i$ respectively. As this is required for  the special case $H=\Lambda$, as the numerator and denominator must cancel. \\

Another informative cross check is given by the  Caley-Hamilton identity, which can be written as $\Pi_{j} (H-\lambda_j I)=0$, therefore
\begin{align}
\Pi_{j} (H-\lambda_j I)|_{\alpha \gamma} &=  (H-\lambda_i I)_{\alpha \beta} ~ \Pi_{j \neq i} (H-\lambda_j I)|_{\beta \gamma} \propto  (H-\lambda_i I)_{\alpha \beta} V_{\beta i}V^*_{\gamma i} =0 \notag \\
&= \Pi_{j \neq i} (H-\lambda_j I)|_{\alpha \beta}   (H-\lambda_i I)_{\beta \gamma} ~ 
\propto   V_{\alpha i}V^*_{\beta i} (H-\lambda_i I)_{\beta \gamma} =0  \,,
\end{align}
as required.

\section{Some Useful Identities:}
\label{sec:SUI}

For n-flavors:
\begin{align}
W_{\alpha \beta}W_{\gamma \delta}-W_{\alpha \delta}W_{\gamma \beta} &= 
\sum_{i>j} m^2_i m^2_j (U_{\alpha i} U_{\gamma j}-U_{\alpha j} U_{\gamma i}) (U_{\beta i} U_{\delta j}-U_{\beta j} U_{\delta i})^* .
\end{align}
For 3-flavors, the unitarity of U gives
\begin{align}
(U_{\alpha i} U_{\gamma j}-U_{\alpha j} U_{\gamma i}) &= 
\text{Det}[U] U^*_{\sigma k}.
\end{align}
where $(\alpha, \gamma, \sigma)$ and (i,j,k) are all different, 
therefore
\begin{align}
W_{\alpha \beta}W_{\gamma \delta}-W_{\alpha \delta}W_{\gamma \beta} &=
\sum_{i>j} m^2_i m^2_j U_{\sigma k} U^*_{\rho k}
\end{align}
where $(\alpha, \gamma, \sigma)$, $(\delta, \beta, \rho)$ and (i,j,k) are all different.

For 4-flavors, the unitarity of U gives
\begin{align}
(U_{\alpha i} U_{\gamma j}-U_{\alpha j} U_{\gamma i}) &= \text{Det}[U]( U_{\beta k} U_{\delta l}-U_{\beta l} U_{\delta k})^* 
\end{align}
where $(\alpha, \gamma, \beta, \delta)$ and (i,j,k,l) are all different.   Therefore
\begin{align}
|U_{\alpha i} U_{\gamma j}-U_{\alpha j} U_{\gamma i}|^2 &= | U_{\beta k} U_{\delta l}-U_{\beta l} U_{\delta k}|^2 
\end{align}
since $ | \, \text{Det}[U] \, |^2 =1$.

\section{Solutions to a cubic equation}
\label{sec:cubic}
The exact solutions to the cubic equation
\begin{align}
\lambda^3-A\lambda^2 +B \lambda -C=0
\end{align}
are given by \cite{Barger:1980tf,Zaglauer:1988gz}
\begin{align}
\lambda_1&=\frac13 A- \frac13 \sqrt{A^2-3B}\left(S+\sqrt3\sqrt{1-S^2} \right)\,, \nonumber \\
\lambda_2&=\frac13 A- \frac13 \sqrt{A^2-3B}\left(S-\sqrt3\sqrt{1-S^2}\right)\,,
\label{eq:3soln_ex} \\
\lambda_3&=\frac13 A+\frac23\sqrt{A^2-3B}\,S\,,  \nonumber 
\end{align}
with
\begin{align}
S& \equiv \cos\left\{\frac13\cos^{-1}\left[\frac{2A^3-9AB+27C}{2(A^2-3B)^{3/2}}\right]\right\}\,,\label{eq:S}
\end{align}
The terms $A$, $B$, and $C$ are the sum of the eigenvalues, the sum of the products of the eigenvalues, and the triple product of the eigenvalues respectively.

As an example of the analytic impenetrability of $S$, eq. \ref{eq:S}, set $A=\sum_i m^2_i$, $B=\sum_{i>j}  m^2_i m^2_j$ and $C=\Pi_i \, m^2_i$, then to recover the eigenvalues, $(m^2_1, m^2_2, m^2_3)$, from eq. \ref{eq:3soln_ex}  is a highly non-trivial exercise.


\section{n-flavor NHS ID}
\label{sec:nNHS}
Assuming all off-diagonal elements of the Hamiltonian in the flavor basis are independent of the matter potential, then
\begin{align}
(2E)^3 ~\Im\biggr[\sum_{\gamma \neq \alpha, \beta} H_{\alpha \beta} H_{\beta \gamma} H_{\gamma \alpha}\biggr] &= 
\Im\biggr[
\sum_{\gamma \neq \alpha, \beta}  \biggr( \sum_i \lambda_i V_{\alpha i} V^*_{\beta i} \biggr) ~
\biggr( \sum_j \lambda_j V_{\beta j} V^*_{\gamma j} \biggr) ~
 \biggr(\sum_k \lambda_k V_{\gamma k} V^*_{\alpha k} \biggr) \biggr]
\end{align}
is an invariant.
Using
$$\sum_{\gamma \neq \alpha, \beta}  V^*_{\gamma j}  V_{\gamma k}  =\delta_{jk} - V^*_{\alpha j}  V_{\alpha k} -V^*_{\beta j}  V_{\beta k}. $$
Only the $\delta_{jk} $ contributes to the imaginary part as the rest lead to real terms. With the notation $J^{\alpha \beta}_{ij}(V) \equiv {\Im}( V_{\alpha i} V^*_{\beta i}   V^*_{\alpha j}  V_{\beta j} ) $, we obtain
\begin{align}
(2E)^3 ~\Im(\sum_{\gamma \neq \alpha, \beta} H_{\alpha \beta} H_{\beta \gamma} H_{\gamma \alpha}) &= 
\sum_{i,j} \lambda_i \lambda^2_j  ~ J^{\alpha \beta}_{ij}(V),  \notag \\
&= \sum_{i>j} (\lambda_i-\lambda_j)(\lambda_j-\lambda_k)(\lambda_k-\lambda_i)  J^{\alpha \beta}_{ij}(V)
\end{align}
for some fixed k, usually k=1.  The last line is derived using the properties of $J^{\alpha \beta}_{ij}$ and a similar procedure as in appendix \ref{sec:CPVterm}. No terms proportional to $J^{\alpha \beta}_{ik}$ or $J^{\alpha \beta}_{kj}$ appear. Therefore,
\begin{align}
 \sum_{i>j} \Delta m^2_{ij}  \Delta m^2_{jk} \Delta m^2_{ki} ~ J^{\alpha \beta}_{ij}(U)
&= \sum_{i>j} \Delta \lambda_{ij}  \Delta \lambda_{jk}   \Delta \lambda_{ki}  ~J^{\alpha \beta}_{ij}(V)
\end{align}
is the NHS identity for arbitrary number of sterile neutrinos. There are (n-1)(n-2)/2 terms on both sides of this identity, none of them contain $J^{\alpha \beta}_{ik}$ or $J^{\alpha \beta}_{kj}$.

\newpage

\bibliography{Diag}

\end{document}